\documentclass [11pt,twoside,a4paper]{article}
\usepackage{shrthnds}
\usepackage{cite}
\usepackage{graphicx}
\usepackage{hyperref}
\usepackage{slashbox}
\usepackage{axodraw4j}
\usepackage[usenames]{color}
\usepackage{subfigure}
\usepackage{setspace}
\usepackage[hang,small]{caption}
\usepackage{amsmath,amssymb}
\usepackage{slashed}   
\usepackage{float} 
\usepackage{placeins}
\restylefloat{table}             
\usepackage{geometry}
\usepackage{fancyhdr}
\newcommand{\auth}[2]{{\large #1}\footnote{email:~#2}}
\newcommand{\affS}[1]{{\it #1}}
\newcommand{\verttle}[2]{\vspace{-1cm}\begin{flushright}{\small #1}\end{flushright}\vspace{0.5cm} {\sf\bfseries #2}}
\newcommand{\nabstract}[2][]{\bc\begin{minipage}{0.9\textwidth}\begin{spacing}{1}{\small {\sf\bfseries Abstract:} #2 }\end{spacing}
#1 \end{minipage}\ec}
\newcommand{\nkeywords}[1][]{~\\\small{ {\sf\bfseries Keywords:} #1 }}

\usepackage{ulem}

\newcommand{\GGHHH}{gg \to HHH}
\newcommand{\GGHHZ}{gg \to HHZ}
\newcommand{\QQHHZ}{qq \to HHZ}
\newcommand{\tin}[1] {\textit{\tiny #1}}
\newcommand{\sbx}[2] {\scalebox{#1}{ #2}}

\usepackage{color}
\definecolor{viv}{RGB}{255,109,254}

\title{\verttle{}{Production of $HHH$ and $HHV(V=\gamma,Z)$	 at the hadron colliders \\}}
\author{
  \auth{Pankaj Agrawal$^{a,b}$}{agrawal@iopb.res.in}~,
  \auth{Debashis Saha$^{a,b}$}{debasaha@iopb.res.in}~, and
  \auth{Ambresh Shivaji$^c$}{ambresh.shivaji@uclouvain.be}\\~\\  
  \affS{a) Institute of Physics, Sainik School Post, Bhubaneswar 751 005, India}\\
  \affS{b) Homi Bhabha National Institute,Training School Complex, }\\
   \affS{ Anushakti Nagar, Mumbai 400085, India}\\
  \affS{c) Centre for Cosmology, Particle Physics and Phenomenology (CP3)} \\
  \affS{ Université Catholique de Louvain, B-1348 Louvain-la-Neuve, Belgium}\\~\\
}
\newcommand{\pghdr}{}
\date{\today}

\begin{document}
\vspace{1.5in}
\maketitle

\nabstract[\nkeywords{Electroweak, Higgs boson, LHC, Anomalous couplings}]{
We consider the production of two Higgs bosons in association with a gauge
boson or another Higgs boson at the hadron colliders. We compute the
cross sections and distributions for the processes  $ p p  \to H H H $ and $H H Z$ within the standard model. 
In particular, we compute the gluon-gluon fusion one-loop contributions mediated via heavy quarks in the loop.
It is the leading order contribution 
to $ p p  \to  H H H $ process. To the process $ p p  \to  H H Z $, it is next-to-next-to-leading-order 
(NNLO) contribution in QCD coupling. 
We also compare this contribution to the next-to-leading-order  (NLO) QCD contribution 
to this process. The NNLO contribution can be similar to NLO contribution
at the Large Hadron Collider (LHC), and significantly more at higher center-of-mass
energy machines. We also study new physics effects in these processes
by considering $ttH, HHH, HHHH, HZZ$, and $HHZZ$  interactions as anomalous.
The anomalous  couplings can enhance the cross sections significantly. The $\GGHHH$ process is specially sensitive to anomalous trilinear Higgs boson self-coupling. For the $\GGHHZ$ process, there is some modest dependence on anomalous $HZZ$ couplings.}

\bigskip

\section{Introduction}\label{sec:introduction}

The CMS and ATLAS Collaborations have been collecting data at the Large Hadron
Collider (LHC) for several years \cite{ATLASRecent, CMSRecent}. Their major 
discovery has been of much-anticipated
Higgs boson in 2012 \cite{ATLASHiggs,CMSHiggs}. There are many reasons to go beyond the standard 
model (SM). Since 2012, the search has been going on for any hint for physics
beyond the standard model. There have been a number of anomalies that
were suspected at different times \cite{ATLASAnomaly1,ATLASAnomaly2,CMSAnomaly1}, 
however none has stood test of times.
There is no evidence of any signal for beyond the standard model scenarios. A number of popular scenarios involving supersymmetry and large extra dimensions
are getting severely constrained \cite{ATLASsusy,CMSsusy,ATLASbsm1,CMSbsm1}.
 Various processes are being analyzed for any hint of a new scenario 
\cite{ATLASRecent, CMSRecent}. In such a situation, exploration of rare processes
and radiative corrections provide promising avenues to explore. We should note
that some other experiments, e. g. LHCb, have reported some unexplained phenomena
\cite{LHCbBKll}.

At a hadron collider, as centre-of-mass energy increases, so does gluon-gluon
luminosity. Therefore, at the LHC and at future probable hadron colliders,
the gluons initiated processes would play important role. In this paper, we are
considering few such processes. The processes that we consider occur at one-loop level. 
The gluon-gluon initiated $2 \to 3$ one-loop processes have been considered in the literature.
First full calculation of $ g g \to \gamma \gamma g$ was presented in \cite{Agrawal:1998ch}.
Many other authors have also computed the contribution of the gluon-gluon
initiated processes on  `multi-bosons +jets'
\cite{deFlorian:1999tp, Melia:2012zg,Agrawal:2012df,Campanario:2011cs,Agrawal:2012as,
Campanario:2012bh,Shivaji:2013cca,Campbell:2014gua,Agrawal:2014tqa,Mao:2009jp,Hespel:2015zea,Hirschi:2015iia,
Gabrielli:2016mdd}.  
{ These processes are important ingredients for the NLO QCD predictions for $gg \to BB(B=\gamma,Z,W,H)$ 
processes~\cite{Caola:2015psa,Caola:2015rqy,Campbell:2016yrh,Caola:2016trd,Granata:2017iod}.}
These different calculations use
different reduction techniques, different packages for computing scalar integrals, and
overall different philosophy for the computation. We use our own tensor-reduction
code, and have developed a comprehensive package for such calculations.

In this paper, we consider the processes  $ g g   \to  H H H, H H \gamma $, and $H H Z$,
and their contribution to hadron level processes -- $ p p  \to  H H H, H H \gamma $,
and $H H Z$. { We presented some preliminary results on these processes along with 
other Higgs processes in $2 \to 3$ category in a conference proceeding~\cite{Shivaji:2016lnu}.}
Triple Higgs production via gluon fusion has been studied by many authors~\cite{PlehnRauch,Binoth:2006ym,Maltoni:2014eza,Papaefstathiou:2015paa,Fuks:2017zkg,Kilian:2017nio}.
Plehn and Rauch \cite{PlehnRauch} considered the possibility of measuring
quartic Higgs boson coupling in the $HHH$ production. In the reference \cite{Maltoni:2014eza},
authors have considered
the QCD correction to the $HH$ and $HHH$ production in a Higgs effective
field theory approach. 
The authors in \cite{Papaefstathiou:2015paa,Fuks:2017zkg,Kilian:2017nio},
have considered the possibility of observing the $HHH$ production at a 100 TeV 
collider including new physics effects. 
{In Ref.\cite{Hirschi:2015iia} standard model cross sections for a number of loop-induced gluon 
fusion processes including $gg \to HHH, HHZ$ are reported. While our work on $pp \to HHH$ has some overlap with these papers as discussed
below, our detailed study of $pp \to HHZ$ process in SM and beyond 
is new and being presented here for the first time in the literature. We have also computed using different tools and looked at different aspects of the processes.}

The exploration of the production of $HHH$ is important, as it is one of
the very few processes where quartic Higgs boson coupling is involved. This
process may allow the {\it direct} measurement of this coupling. With the measurement
of self-couplings of the Higgs boson, one can confirm the form of the Higgs
potential.  
Unlike the $HHH$ production,
 the process $ p p  \to  H H Z$ gets contribution from the tree-level
processes. One can also compute next-to-leading order (NLO) QCD corrections
to this tree-level process~\cite{Frederix:2014hta}. In this paper, our focus is on gluon-gluon annihilation
contribution, but we also compare it with the LO and NLO contributions.
The $\GGHHZ$ contribution can be thought of as next-to-next-to-leading order (NNLO)
corrections~\cite{Baglio:2012np}. The production of $HHZ$ is important, as it involves $HHH$ and $HHZZ$ couplings. 
{ It is also a background to triple Higgs production process.}

In search for  new physics scenarios, the use of anomalous
interactions may play an important role. This is a model-independent approach
which can systematize the search. They may point towards a model
that would be suitable to go beyond the standard model. In this paper,
we consider possible modification of standard model interactions, inspired
by dimension-six operators in effective field theory approach. In particular,
we consider the modifications of  $ttH, HHH, HHHH, HZZ$, and $HHZZ$ interactions.
{We study the effects of these anomalous interactions on the production cross 
sections and on kinematic distributions in $gg \to HHH, HHZ$ processes.} 
{Note that in Refs.~\cite{Papaefstathiou:2015paa,Fuks:2017zkg}, the $HHH$ process is studied 
considering only SM-like deviations in trilinear and quartic couplings. 
We have in addition considered derivative couplings which have different
effects on some distributions as we demonstrate.
In Ref.~\cite{Kilian:2017nio}, all the CP-even dimension-six operators
 relevant to $HHH$ process has been considered. In present work, our approach towards 
 new physical effects in $HHH$ process is more phenomenological and we have considered modifications to 
 only those couplings which are present in the standard model at tree-level. 
 For example, we do not consider $ttHH$, $ttHHH$, $ggH$ and $ggHH$ interactions. However, 
 for $ttH$ coupling we have considered both CP-even and CP-odd anomalous interactions.  
 Similar approach is taken to study new physics effects in $HHZ$ process. }

The paper is organized as follows. In the next section~\ref{sec:processes}, we discuss these
various processes and gluon-gluon annihilation contributions to them.
In the section~\ref{sec:anml}, the anomalous contribution to various relevant vertices
is discussed.  In the section~\ref{sec:calculation}, we provide details on the 
 method of calculation and numerous checks. 
Our numerical results for SM and BSM are presented in section~\ref{sec:results}. 
 In the last section~\ref{sec:conclusions}, we present our conclusions.

\section{Processes}\label{sec:processes}

 We consider the gluon-gluon annihilation contribution to the
 following processes:
\begin{eqnarray}
 p ~p &  \to &  H ~H ~ \gamma, \\
 p ~p & \to &  H ~H ~H , \label{eq:thj}\\
 p ~p & \to &  H ~H ~Z . \label{eq:thw}
\end{eqnarray}

Let us first consider the process $ p ~p \to H ~H ~\gamma$. In the standard model, at tree-level,
this process
has vanishingly small cross section due to very small light-quark and Higgs boson coupling. At one-loop,
there is no contribution to this process form gluon-gluon fusion channel. Using Furry's theorem, one can see that
all diagrams contributing to this process add up to zero.

 \begin{figure}[h]
\bc
\includegraphics [angle=0,width=.75\linewidth]{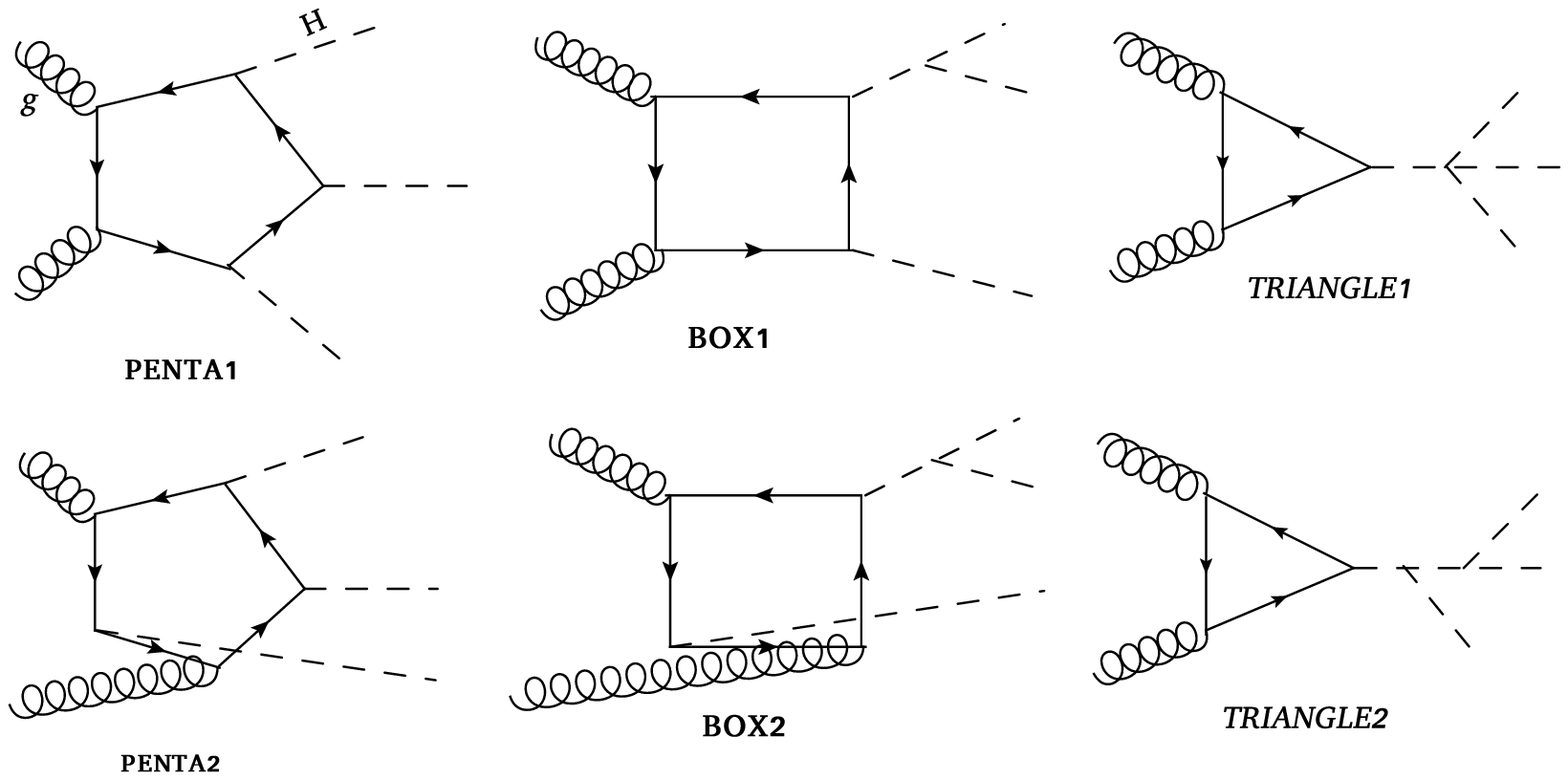}
\ec
\caption{Different classes of diagrams contributing to $\GGHHH$ process. } \label{fig:diagrams-hhh}
\end{figure}

The  $ p ~p  \to   H ~H ~H $ process occurs primarily via $ g ~g  \to  H ~H ~H$ and leading contribution comes from 
one-loop diagrams. Because of the very small light-quark and Higgs boson couplings, the tree level process
$q {\bar q} \to H H H$ makes very small contribution. Therefore, we only consider the contribution of the
$ g ~g  \to  H ~H ~H$ process. This is a one-loop process. There are 24 pentagon, 18 box, and 8 triangle 
diagrams contributing to the process for each quark flavor in the loop. There are some more  diagrams 
in the box and triangle categories at one-loop level, but their contributions are zero, as they do not 
conserve color charge (Tr($\lambda^a$)=0). In the loop, a light-quark would not contribute due to very 
small Yukawa coupling; so we keep only diagrams with a top-quark and a bottom quark. 
We do not need to numerically compute all the diagrams separately, as many diagrams are related to one another by charge conjugation 
symmetry, or crossing. We can compute all the diagrams using six prototype diagrams as shown in Fig.~\ref{fig:diagrams-hhh}. 
By permuting the legs in the prototype diagrams, and using charge conjugation (Furry's theorem), all the 
amplitudes can be calculated. Out of 24 (=4!) pentagon diagrams, we had to numerically calculate only 12 
diagrams as other 12 diagrams can be related to the previous diagrams by Furry's theorem. Out of these 12 
pentagon diagrams, 6 (=3!) diagrams can be obtained from PENTA1 prototype diagram by permuting 
the Higgs bosons; other 6 diagrams can be obtained similarly from  PENTA2 prototype diagram. Similarly, out 
of 18 box diagrams, we need to numerically compute only 9 diagrams, and from these the rest can be found out using Furry's theorem only. Out of these 9 box diagrams, 6 can be calculated from BOX1 prototype diagram by permuting Higgs bosons in six different ways. 
 Other 3 box diagrams can be obtained by the permutation of external Higgs boson  in BOX2 prototype 
diagram. Out of 8 triangle diagrams, we need to numerically compute only 4; TRIANGLE1 prototype diagram gives one of these and 
TRIANGLE2 prototype diagram gives the rest 3 
diagrams by permuting external Higgs bosons. 

\begin{figure}[h]
\bc
\includegraphics [angle=0,width=.75\linewidth]{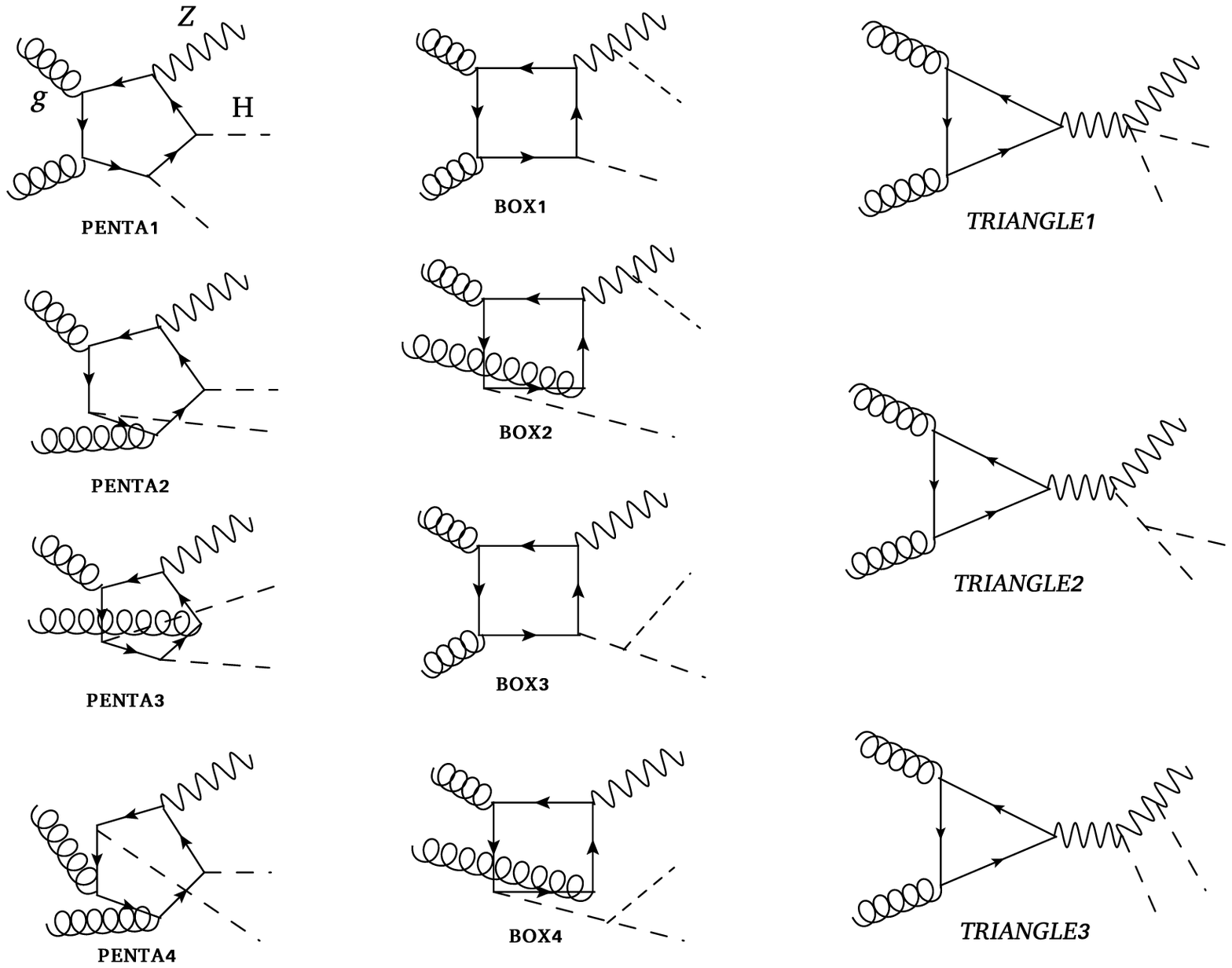}
\ec
\caption{Different classes of diagrams contributing to $\GGHHZ$ process. } \label{fig:diagrams-hhz}
\end{figure}

The  $ p ~p  \to  H ~H ~Z $ can occur at tree level through quark-antiquark annihilation. 
The cross section is large enough for it to be observable with high luminosity option.
One can also compute NLO QCD corrections
to this process easily. Our focus will be contribution of  gluon-gluon fusion process: $ g ~g  \to  H~H ~Z $ 
which occurs at the one-loop level. Formally this contribution is of NNLO order. However, 
as we shall see, at the LHC,  this contribution
can be of the same order as the NLO corrections.  Due to enhanced gluon-gluon luminosity at larger
center-of-mass energies, this NNLO correction will even dominate over NLO correction at higher energy
hadron colliders.  
There are 24 pentagon, 18 box, and 8 triangle diagrams contributing to the process $ g ~g  \to  H ~H~Z $. 
Like in the case $ g ~g  \to  H~H~H $, here also the diagrams are divided into many prototype classes as 
shown in Fig.~\ref{fig:diagrams-hhz} . Here, as before,  out of 24 pentagon, 18 box, and 8 triangle diagrams, we need to numerically compute only 12 pentagon, 
9 box, and 4 triangle diagrams respectively. Out of these 12 pentagon diagrams,  
each of PENTA1 and PENTA2  prototype diagrams gives four diagrams, and each of PENTA3 and PENTA4  
prototype diagrams gives two diagrams. In the case of box diagrams, BOX1, BOX2, BOX3, and BOX4 
prototype diagrams give 4, 2, 2, and 1 diagrams respectively. There are three prototype triangle diagrams -- TRIANGLE1, TRIANGLE2, and TRIANGLE3; these prototype diagrams 
give 1, 1, and 2 diagrams respectively. Note that due to Furry's theorem only the 
axial-vector part of the $Z$ boson coupling with the quarks in the loop contributes. Feynman diagrams in Fig.~\ref{fig:diagrams-hhh} and Fig.~\ref{fig:diagrams-hhz} have been made using JaxoDraw~\cite{Binosi:2008ig}.
Computation of one-loop diagrams in $gg \to HHH, HHZ$ processes are described in section 4.


\FloatBarrier

\section{Anomalous interactions of Higgs boson}\label{sec:anml}

There are a number of arguments for going beyond the standard model.
In absence of any new resonance at the LHC, one way to consider the effects of beyond-the-standard-model scenarios
is to consider the possible modification of standard  model vertices. 
{We are mainly interested in anomalous couplings of the Higgs boson 
which would affect the processes under consideration. These include, 
$ttH, HZZ, HHZZ, HHH$ and $HHHH$ interactions. Some of them, for example, 
$ttH$ and $HZZ$ are already constrained by the existing LHC data~\cite{PDG}. 
The trilinear Higgs self coupling is very weekly constrained by the data~\cite{Sirunyan:2017djm} 
and couplings $HHZZ$ and $HHHH$ are unconstrained at present. In the following, 
we consider most general interaction Lagrangian incorporating the BSM physics 
which would lead to deviations in the Higgs couplings of our interests.}

\subsection{Anomalous $\bar{t}tH$ Vertex}\label{sec:anml-tth}

In the standard model, the top quark couples with the Higgs boson via
 the Yukawa coupling. This leads to a scalar $ttH$ coupling.  
The most general vertex for $\bar{t} t H$ interaction can be parametrized as,
\ba
\mc L_{\scriptscriptstyle {{\bar t}tH}} = -\frac{m_\tin{t}}{v}\bar t\left[\left(1+ \kappa_\tin{t}\right)  + i {\tilde \kappa}_\tin{t} \g_5\right]t H.\label{eq:tth_lag}
\ea
 Here $v$ is electroweak symmetry breaking scale and it is approximately $247$ GeV.  
In the standard model, ${\kappa}_\tin{t} = {\tilde \kappa}_\tin{t} = 0$. 
We use following bounds for ${\kappa}_\tin{t}$ and ${\tilde \kappa}_\tin{t}$ 
\cite{Whisnant:1994fh, Nishiwaki:2013cma}:
\begin{eqnarray}
&& -0.2 \leq {\kappa}_\tin{t} \leq  0.2\;,  \nonumber\\
&& -0.1 \leq {\tilde \kappa}_\tin{t} \leq  0.1\;.\label{blindlim}
\end{eqnarray}

This vertex contributes to both $HHH$ and $HHZ$ production. 
As we shall see, the scaling of the scalar coupling, as parametrized by 
${\kappa}_\tin{t}$ can change the cross section significantly.

\subsection{Anomalous $HHH$ and $HHHH$ Vertices}\label{sec:anml:hhh}

After the discovery of the Higgs boson, one of the important task is to determine the 
form of the Higgs potential. As discussed before, one of the characteristic feature of the 
Higgs potential in the standard model is specific form of
Higgs boson self-couplings. 
The Higgs boson self-interactions can be expressed in terms of the anomalous couplings as
\begin{eqnarray}
    \label{Hstext}
    \mathcal{L}_\tin{HHH} &=&
        -{3 m_\tin{H}^2\over v}
        \left(\dfrac{1}{6} (1+ g^{(0)}_{\text{\tiny 3}\tin{H}})\; H^3 + \frac{1}{6 m_\tin{H}^2} g^{(1)}_{\text{\tiny 3}\tin{H}}\; {H \partial_\mu H \partial^\mu H} \right),\\
    \mathcal{L}_\tin{HHHH} &=&
        -{ 3 m_\tin{H}^2\over v^2} \left(\dfrac{1}{24}(1+ g^{(0)}_{\text{\tiny 4}\tin{H}})\; H^4 +
      \frac{1}{24 m_\tin{H}^2} g^{(1)}_{\text{\tiny 4}\tin{H}}\;  {H^2\partial_\mu H \partial^\mu H} \right).
\end{eqnarray}

In the SM, $g^{(0)}_{\text{\tiny 3}\tin{H}}=g^{(0)}_{\text{\tiny 4}\tin{H}}=g^{(1)}_{\text{\tiny 3}\tin{H}}=
g^{(1)}_{\text{\tiny 4}\tin{H}}=0$.
Here $g^{(0)}_{\text{\tiny 3}\tin{H}}$ and $g^{(0)}_{\text{\tiny 4}\tin{H}}$ just scale the
trilinear and quartic Higgs boson self-couplings, respectively. 
Both of these couplings occur in $HHH$ production. In the $HHZ$ production, only trilinear
coupling occurs. As we shall see, both the processes are sensitive to modification of
trilinear coupling.
These couplings are poorly determined, as they occur in processes with small cross section.
It may take a decade or more before a serious bound can be put on these couplings.
For illustration, we take these parameters in the range between -1.0 to 1.0.

\subsection{Anomalous $HZZ$ and $HHZZ$ vertices }\label{sec:anml-hzz}

  These two vertices occur in the process $\GGHHZ$. While $HZZ$ vertex occurs in other
 processes, like $p p \to HZ$, the vertex $HHZZ$ occurs mainly in processes involving
 double Higgs boson production, so is poorly constrained. 
   The most general $HZZ$ interaction 
   that can be written has following form,

\ba
{\cal L}_{\tin{HZZ}} &=& \dfrac{g M_\tin{Z}}{c_\tin{W}} \;
\Big\{\dfrac{1}{2}(1+g^{(0)}_\tin{HZZ}) H Z_{\mu} Z^ {\mu}
 -\dfrac{1}{4 M_\tin{Z}^2 } g^{(1)}_\tin{HZZ} {H Z_{\mu \nu} Z^{\mu \nu}} -
  \dfrac{1}{ M_\tin{Z}^2 } g^{(2)}_\tin{HZZ}\ {H Z_{\nu} \partial_{\mu} Z^{\mu \nu}} \Big\}  \;.
\label{H} 
\ea

In the above $g$ is the coupling parameter of the $SU(2)_L$ group and $c_W = {\rm cos} \theta_W$, $\theta_W$ 
being the weak angle. The SM $HZZ$ interaction corresponds to 
$g^{(0)}_\tin{HZZ} = g^{(1)}_\tin{HZZ} = g^{(2)}_\tin{HZZ} = 0$.
 This interaction has derivative couplings which lead to momentum dependence. This is unlike 
 standard model.

The following are the various bounds on the anomalous coupling parameters \cite{Ellis:2014}
\begin{eqnarray}
&& -0.10\leq \ g^{(0)}_\tin{HZZ} \leq  0.10\;,  \nonumber\\
&& -0.09\leq \ g^{(1)}_\tin{HZZ} \leq  0.04\;,  \nonumber\\
&& -0.07\leq \ g^{(2)}_\tin{HZZ} \leq  0.03\;.  \label{blindlim}
\end{eqnarray}
There is also modification of $HHZZ$ coupling. This modification only scales the
standard model interactions:
\ba
{\cal L}_{\tin{HHZZ}}&=& \dfrac{g M_\tin{Z}}{c_\tin{W} v} \;
\Big\{\dfrac{1}{4}(1+g^{(0)}_\tin{HHZZ})\; H H Z_{\mu} Z^ {\mu} \Big\}  \;.
\label{H} 
\ea

{In the SM, $g^{(0)}_\tin{HHZZ}=0$. In absence of any available bound for this coupling, we allow the parameter $g^{(0)}_\tin{HHZZ}$
to vary between -0.1 and 0.1.

The anomalous interactions mentioned above are well motivated within the framework 
of an effective field theory in which the new physics effects are parametrized in 
terms of higher dimensional operators. These operators are constructed from the 
SM fields and respect the symmetries of the SM. 
A complete list of independent dimension-six operators is now available~\cite{BasisBW, BasisWarsaw, BasisSILH}. 
The anomalous couplings introduced above are related to the Wilson coefficients of these 
operators~\cite{Saavedra1,Barger:2003rs,Gonzalez:99,Hagiwara:1993ck, Zhang:2003it}.

The Feynman rules for the anomalous Higgs vertices are listed in the appendix~\ref{sec:A-FR}.
As we shall see, in the allowed range of the parameter values, 
the contribution of the anomalous vertices can be important in our processes. 
Note that even in the presence of these anomalous couplings the 
amplitude for $ gg \to HH\gamma$ process does not receive any contribution. 
}

\section{Calculation and Checks}\label{sec:calculation}
   
      As discussed in the section 2, we compute prototype pentagon, box, and triangle 
      diagrams. Then by using crossing and Furry's theorem, we can compute rest of
      the diagrams. All the diagrams have a fermion loop, so to calculate the amplitude
      we need to compute  trace  of a string of gamma matrices.
      We calculate the traces of the prototypes diagrams using  {\tt FORM} \cite{Vermaseren:2000nd}.  
     The process $ g g \to HHZ$ includes $ttZ$ coupling, which has both vector and axial vector
     parts. The presence of $\gamma_5$ requires special care, due to the potential presence 
     of anomalies.  We have handled this situation in two different ways.  
     Since the process is free from UV divergences, we can take trace in four dimensions. 
     We have also calculated the trace in $n$ dimensions 
     using Larin's prescription for $\gamma_5$~\cite{larin,Shivaji:2011ww}. 
     Both methods, in the end give same results when the contributions from 
     both the top and bottom quark loops are considered. When we include $ttH$ anomalous pseudo-scalar coupling, the trace will include more $\gamma_5$ matrices for both $HHH$ and $HHZ$ production.
   
     In the first step, we use {\tt FORM} to take the trace and
     write the amplitude in terms of tensor and scalar integrals. We use an in-house package {\tt OVReduce} 
     to reduce tensor integrals to lower-point tensor integrals and scalar integrals {in dimensional 
     regularization} . This package is based 
      on the methods of Oldenborgh and Vermaseren \cite{vanOldenborgh:1989wn,Shivaji:2013cca}.  After the reduction, 
     all that we need to do is  to compute various scalar integrals of box, triangle, and bubble types. 
     To compute these scalar integrals, we use {\tt OneLOop} library by 
     Andreas van Hameren \cite{vanHameren:2010cp}. It uses dimensional regularization to regulate UV 
     and IR divergences. For the pentagon scalar integrals, we use van Neerven-Vermaseren technique \cite{vanNeerven:1983vr,Shivaji:2013cca}.
      As the amplitudes are quite large  and complicated, we first compute
      helicity amplitude for a process numerically before squaring it.
      The Monte Carlo integration over the entire phase space has been performed using {\tt VEGAS} code
       \cite{Vegas:1980} as   implemented in {\tt AMCI}   (Advanced Monte Carlo Integration) package~\cite{Veseli:1997hr} .
       {\tt AMCI}  implements a parallel version of {\tt VEGAS} algorithm which makes use of 
       Parallel Virtual Machine ({\tt PVM}) software \cite{pvm}. 
       
       While performing multi-leg one loop calculations, the
        issue of numerical instability comes in for certain phase space points. As the number of 
        such points are not large, and contribution of these phase space points is not expected 
        to be large, as is the practice, we exclude these phase space points by setting a suitable upper bound on the amplitude-squared\footnote{We have 
         cross-checked the robustness of this procedure by comparing it with another implementation of rejecting unstable 
         phase space points based on Ward-identity or gauge invariance check for each phase space point~\cite{Shivaji:2013cca}.}. 
        This upper bound is chosen after finding out possible values the amplitude-squared
        can have by running the code. This upper bound is increased until we hit the
        unstable phase space point. The cross section remains stable and does not change, even when this upper bound
        is increased by several orders of magnitude. For the $HHH$ process, 
        we do not come across any unstable phase space point. For the $gg \to HHZ$ process, 
        the number of unstable points are well below 0.002\%.

We have performed many checks to verify the correctness of the amplitude for each process.
These checks include verification of cancellation of UV $\&$  IR divergences and gauge invariance.
Because of the presence of the Higgs boson in the final state, we have only top and bottom quarks 
in the loop. In both the processes, all pentagon, box, and triangle diagrams  are therefore 
separately IR finite. Overall amplitude at any phase space point is UV finite. Each pentagon diagram is 
UV finite, which is again as expected  as simple power counting
reveals it. Individual box diagrams are also 
UV finite, since one will have at most two (three)-tensor box integrals in $gg \to HHH(HHZ)$.  
Each triangle diagram is also UV finite. 
In both these processes, we 
have verified that in the large top quark mass limit the amplitude 
becomes constant implying non-decoupling of top quark in $m_t \to \infty$ limit~\cite{Appelquist:1974tg}. 

To check gauge invariance, we replace the polarization 
vector of a gluon with its momenta in the amplitude.
In $gg \to HHH$, the overall amplitude has been checked to be gauge invariant with respect to 
both the gluons. We find that each triangle diagram is individually 
gauge invariant, while each pentagon and box diagram is not. However, all 
pentagon diagrams together, and all box diagrams together are
gauge invariant.  Here interesting point is that the pentagon, 
box, and triangle diagrams are separately gauge invariant with respect to the gluons\footnote{ With respect to  
SM EW symmetry only full amplitude is meaningful.}. So it may be tempting 
to use only one class of diagrams to compute the cross section. However as 
will see below, it can lead to serious errors. Here we have done the calculation only in four dimension. The amplitude is found to be gauge invariant for both gluons in the presence of pseudo-scalar coupling even when we consider only one quark.

    We have calculated the $gg \to HHZ$ amplitude treating $\gamma_5$
     in four-dimension and in $n$-dimension using 
    Larin's prescription~\cite{larin}. In four-dimension, if we consider only one quark, each triangle diagram is gauge invariant with respect to one gluon but not for the other. All the triangle diagrams taken together are also not gauge invariant for the other gluon. However, if we consider both top and bottom quarks in the amplitude, 
  each triangle diagram is gauge invariant for both gluons. This is related to the quantum anomaly 
    associated with the axial vector coupling of $Z$ boson with fermions. None of the pentagon or  box diagram is 
individually gauge invariant for any gluon. However, referring to Fig.~\ref{fig:diagrams-hhz}, all the pentagon diagrams taken together, 
or all the diagrams of BOX1 and BOX2 classes together, or all the diagrams of BOX3 and BOX4 classes
 together are gauge invariant for both gluons.
      In $n$-dimension, all the pentagon 
    diagrams together, all the BOX1 and BOX2 diagrams together, all the BOX3 and BOX4 diagrams together, and each triangle diagram 
    are gauge invariant with respect to the both gluons for top and bottom quarks separately. 
    {In any case, to remove the anomaly associated with the chiral current of $Z$ boson 
    the contributions from both the top and bottom quarks in the loop must be 
    included.}

\section{Numerical results}\label{sec:results} 

In our computation of cross sections and distributions  
we have  used: $p_T^{H,Z} > 1 \;{\rm GeV}, |y^{H,Z}| < 5$.
The cut on the $p_T$ of the Higgs boson and the $Z$ boson (of 1 GeV) is there to reduce the number of
 phase space points  which introduce numerical instability. We have checked and it can be 
 understood from the $p_T$ distributions presented below that the effect of removing the cut is negligible. The results for gluon fusion processes are obtained using {\tt cteq6l1} parton 
distribution functions \cite{Nadolsky:2008zw}, and using $\mu_R=\mu_F=\sqrt{\hat{s}}$ 
(partonic center-of-mass energy ) as
the renormalization and factorization scales. We have also included uncertainties 
in the results for the LHC by varying the renormalization and factorization scales 
by a factor of 2. The scale
uncertainties are listed in the table\footnote{With the CT14, the latest version of CTEQ
parton distributions, the cross section for the process $pp \to HHH$ changes by about
$13\%$ at 13 TeV center-of-mass-energy and by about $4\%$ at 100 TeV center-of-mass-energy.}.
 
\subsection{ The process $ pp \to HHH$}\label{sec:results-hhh}

In Table~\ref{table:xs-hhh}, we present the cross section for this process at the LHC center-of-mass energy, and at other proposed hadronic colliders. 
 The cross section at 13 TeV center-of-mass energy
 is 32.0 attobarn. So as of now only 2-3 such events may have  been
 produced at the LHC. With even expected 3 ab$^{-1}$ luminosity, there will be 
 only about 100 events.  At 100 TeV, thanks to a larger gluon flux, the cross section will be 100 times
 larger. Note that these cross sections suffer from large scale uncertainty (-22\% to 31\% at 13 TeV) which is typical to gluon fusion processes.
 {These uncertainties are large due to the significant dependence of the strong coupling constant, $\alpha_{s}(\mu)$, on the renormalization scale. As center of mass energy increases, the coupling constant value decreases, so does the dependence of
 the cross section on the renormalization scale.}

     \begin{table}[h]
 \begin{center}
  \begin{tabular}{|c|c|c|c|c|c}
   \hline
   $\sqrt{\rm s}\;[\rm TeV]$& $8$& $13$& $33$& $100$ \\
   \hline
   \hline
   & & & &\\
     $\sigma^{\textit{\rm \tiny HHH,\,LO}}_{\textit{\rm \tiny GG}}\;~[\rm ab]$
     & $7.0^{+34.6\%}_{-24.0\%}$ & $32.0^{+30.6\%}_{-22.2\%}$ & $330.8^{+23.8\%}_{-18.4\%}$ & $3121.1^{+17.4\%}_{-14.1\%}$ \\
   & & & &\\
   \hline
  \end{tabular}
 \end{center}
 \caption{ $pp \to HHH$ hadronic cross sections and corresponding scale uncertainties in the SM at different collider center-of-mass energies.} 
 \label{table:xs-hhh}
\end{table}


\begin{table}[h]
 \begin{center}
  \begin{tabular}{|c|c||c||c||c|}
   \hline
   {$\sqrt{\rm s}\;[\rm TeV]$} & 8 & 13 & 33 & 100 \\[3pt]
   \hline
   \hline
   & & & &\\
   $\sigma^{\rm \tiny HHH}_{\rm \tiny penta} \; [\rm ab]$ &  22.1 & 94.4 & 916.4 & 8067.8  \\
   & & & &\\
\hline
& & & &\\
   $\sigma^{\rm \tiny HHH}_{\rm \tiny box} \; [\rm ab]$ & 12.9  & 53.6 & 502.5  & 4287.4  \\
   & & & &\\
\hline
& & & &\\
   $\sigma^{\rm \tiny HHH}_{\rm \tiny triangle} \; [\rm ab]$ & 0.8  & 3.5  & 32.1 & 270.8  \\
   & & & &\\
\hline \hline
& & & &\\
   $\sigma^{\rm \tiny HHH}_{\rm \tiny total} \; [\rm ab]$ & 7.0 & 32.0 & 330.3  & 3121.3\\
   & & & &\\
\hline
  \end{tabular}
 \end{center}
  \caption{ SM contribution of pentagon, box, and triangle diagrams to the total cross section in $\GGHHH$ 
  at different collider center-of-mass energies, displaying a destructive interference effect. }
  \label{table:xs-hhh-pen-bx-tr}
\end{table}

\begin{figure}[h]
\bc
\includegraphics [angle=0,width=0.45\linewidth]{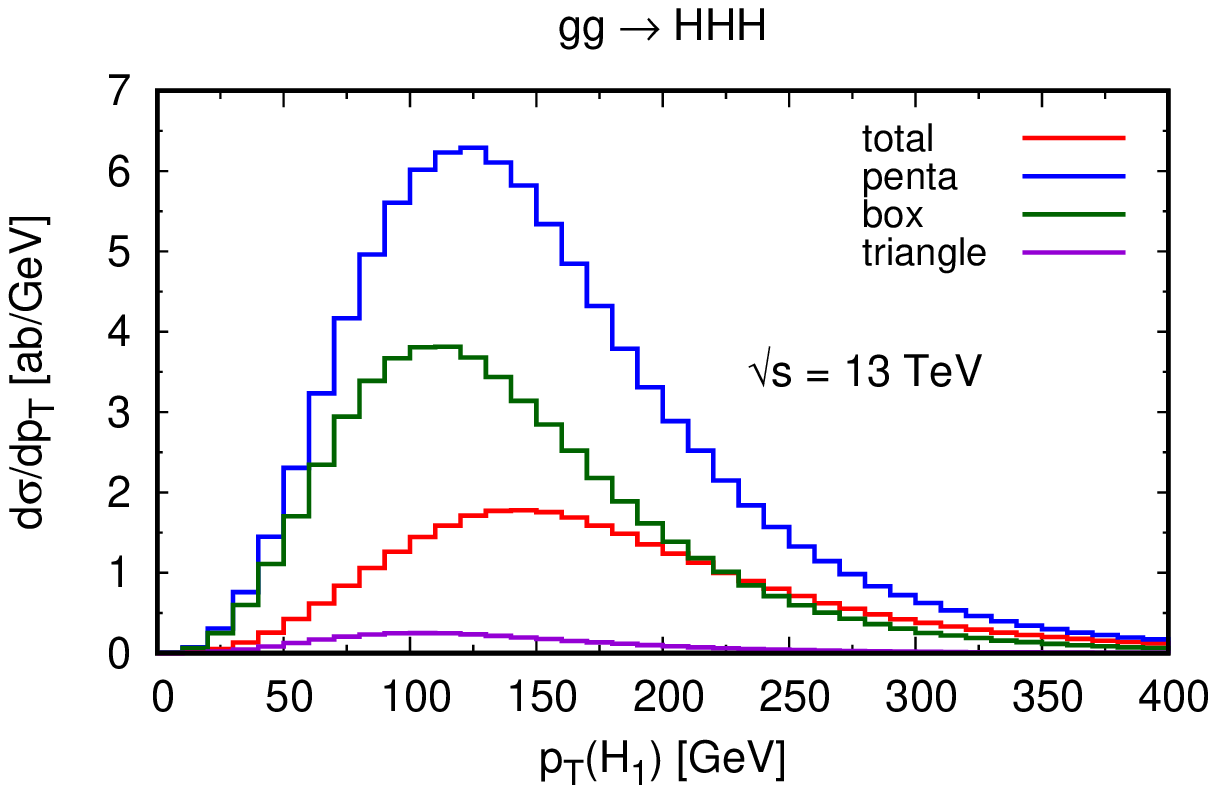}
\includegraphics [angle=0,width=0.45\linewidth]{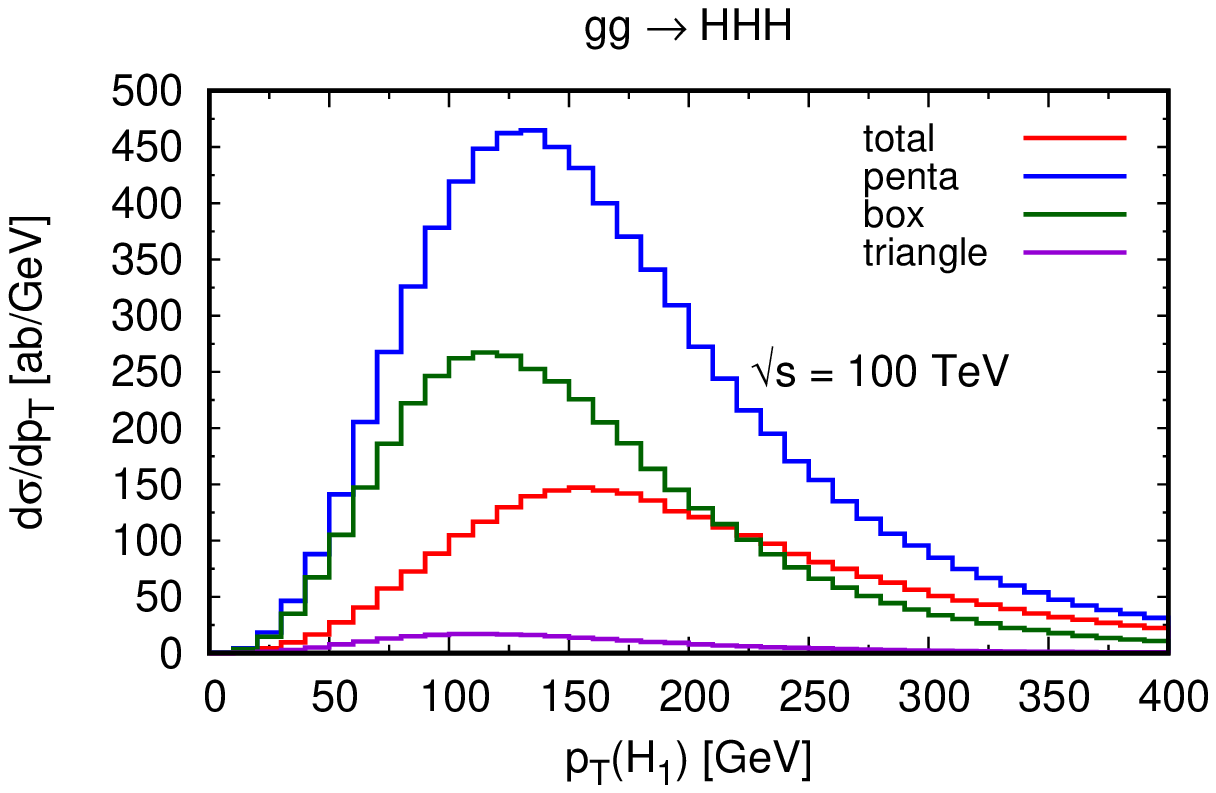}
\ec
\caption{SM contribution of pentagon (blue), box(green), and triangle (violet) diagrams to 
leading  $p_T(H)$ distribution in $\GGHHH$ at 13 TeV (left) and 100 TeV (right).  
} \label{fig:dist-hhh-pen-bx-tr}
\end{figure}

\begin{figure}[!hbt]
\bc
\includegraphics [angle=0,width=0.45\linewidth]{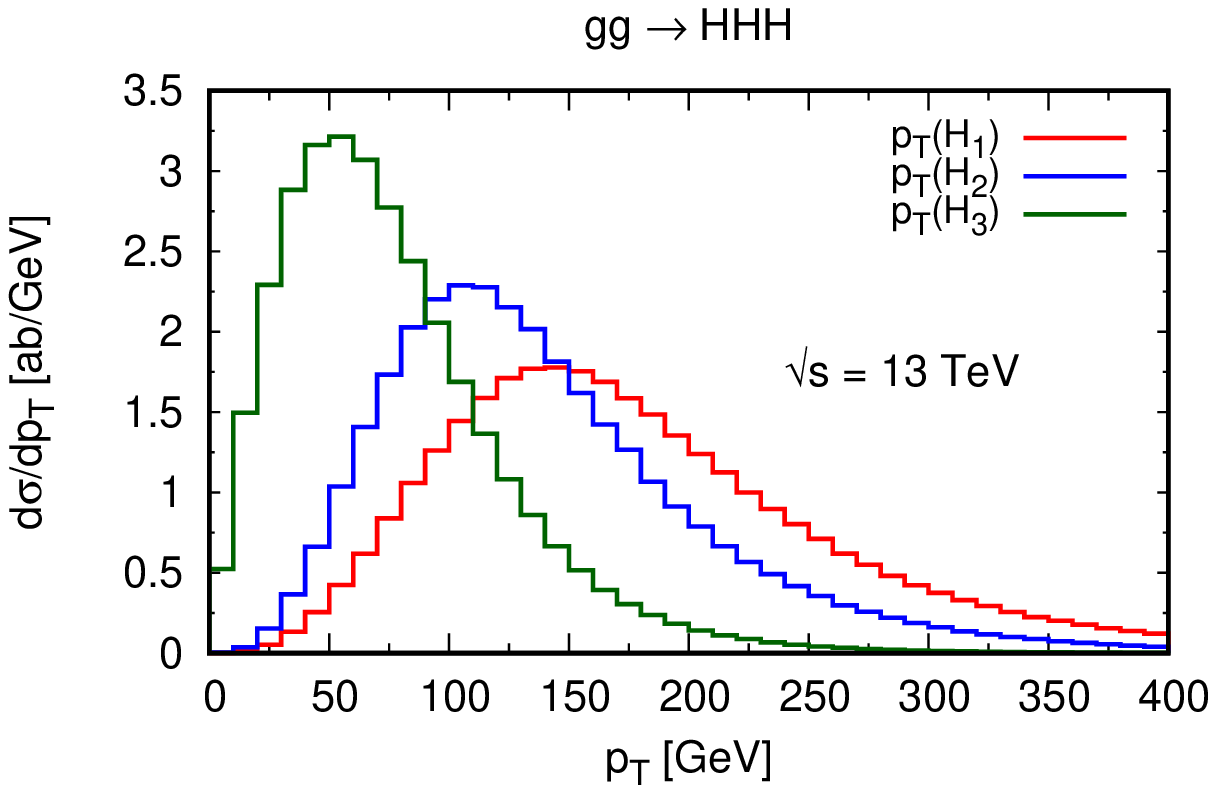}
\includegraphics [angle=0,width=0.45\linewidth]{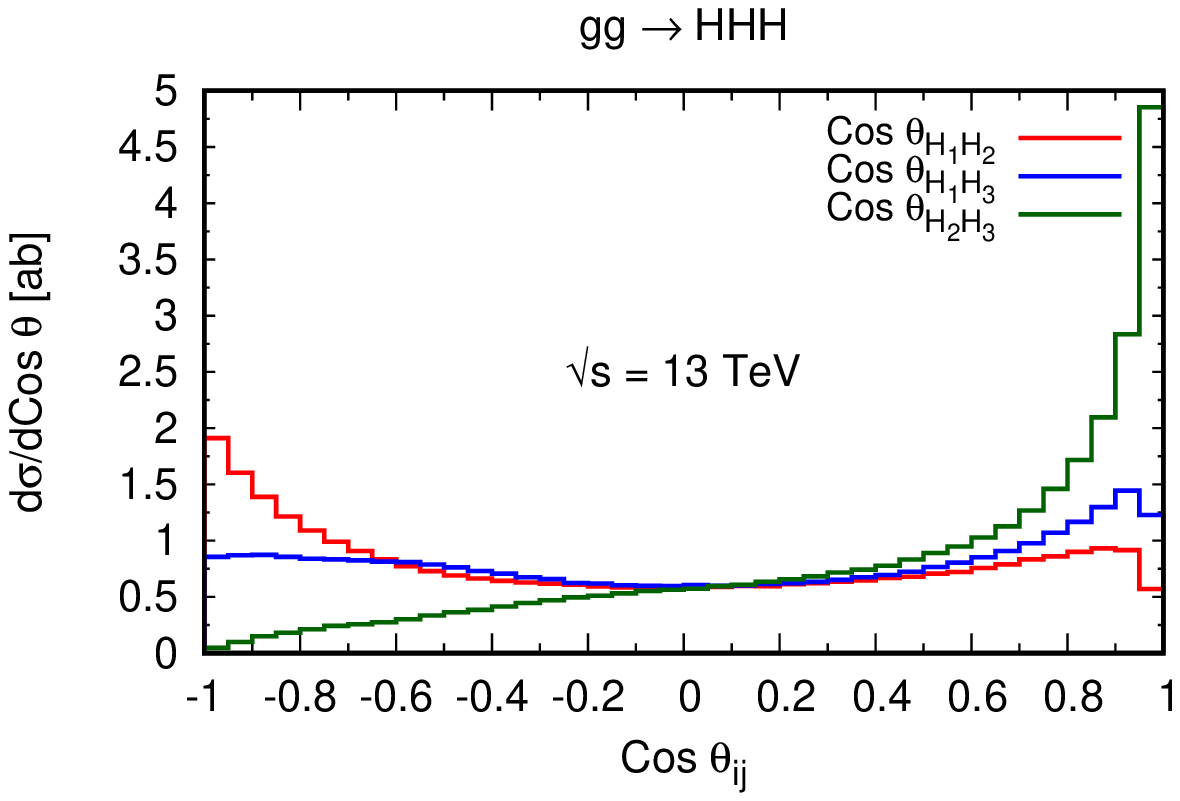}
\includegraphics [angle=0,width=0.45\linewidth]{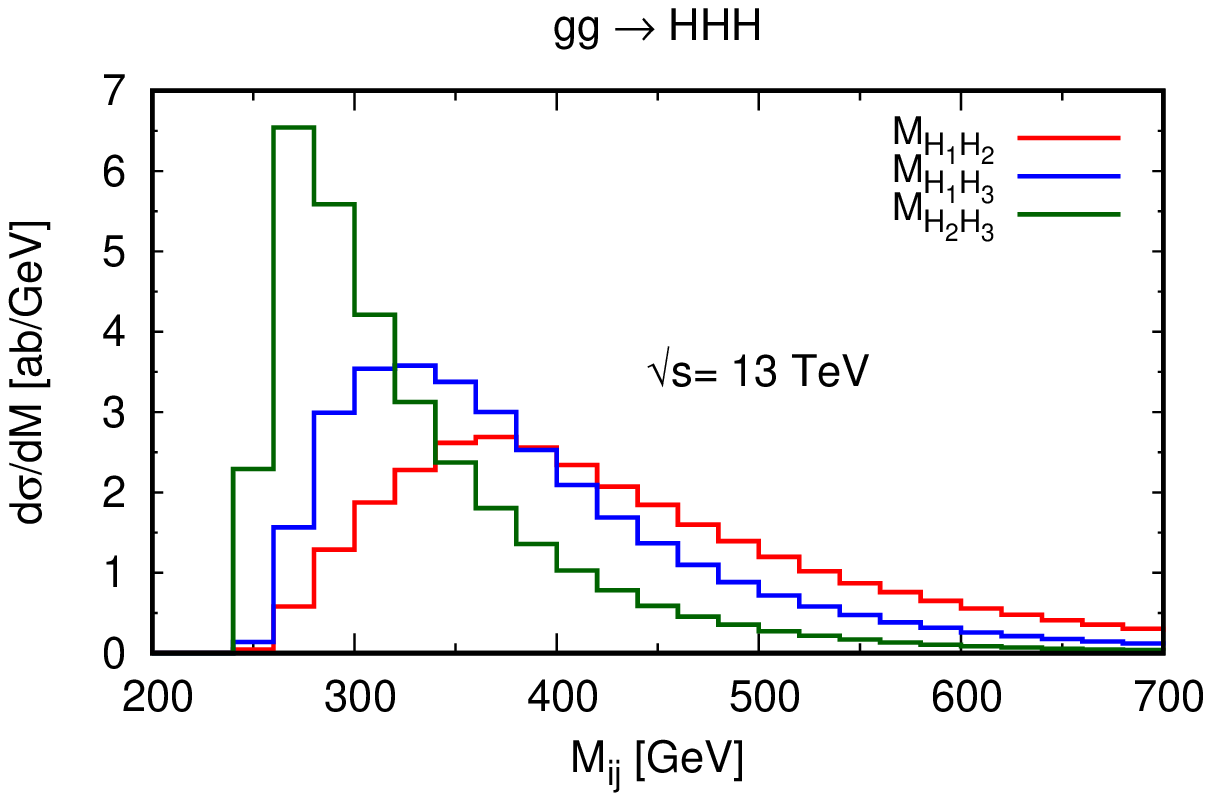}
\includegraphics [angle=0,width=0.45\linewidth]{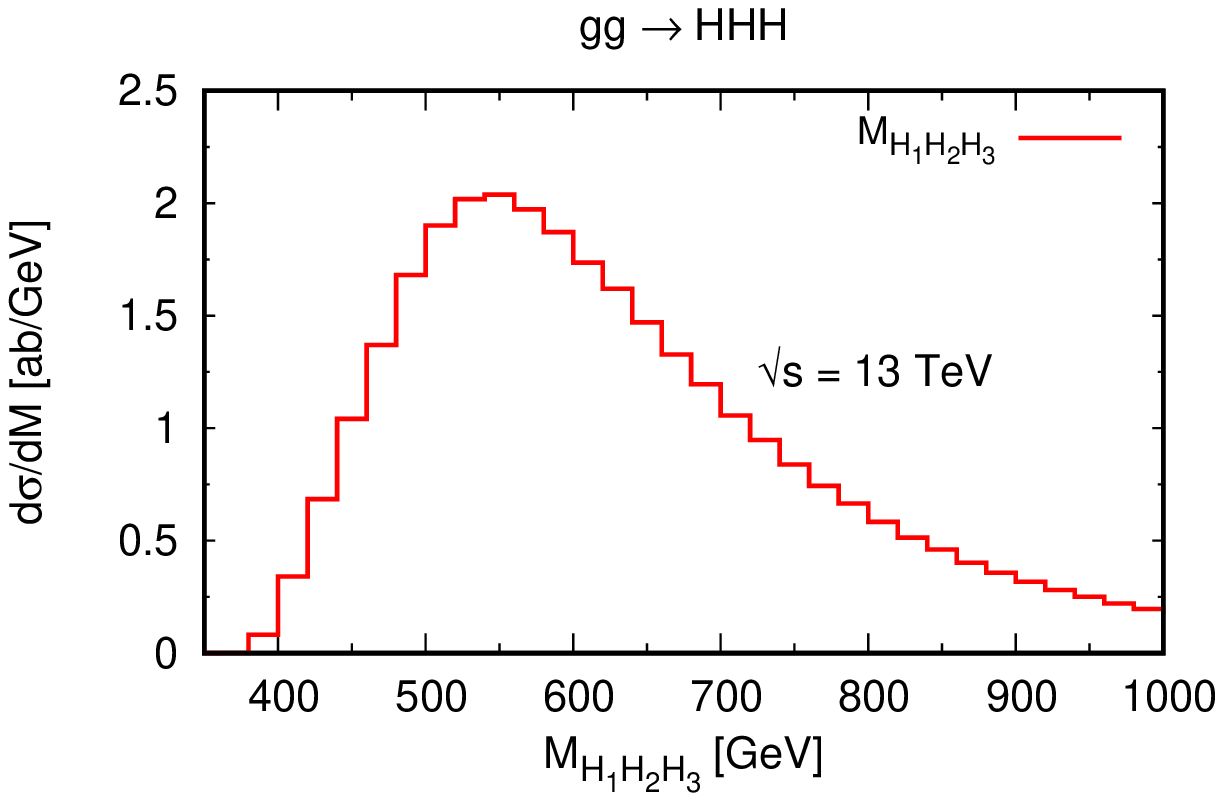}
\ec
\caption{ Kinematic distributions for $\GGHHH$ in the SM at 13 TeV. These plots are obtained 
after $p_T$ ordering the Higgs bosons. $H_1, H_2$, and $H_3$ refer to the 
hardest, second hardest, and third hardest Higgs bosons in $p_T$ respectively.} 
\label{fig:dists-hhh-sm}
\end{figure}

 \begin{center}

\begin{figure}[!hbt]
 \begin{center}
 \includegraphics [angle=0,width=0.45\linewidth]{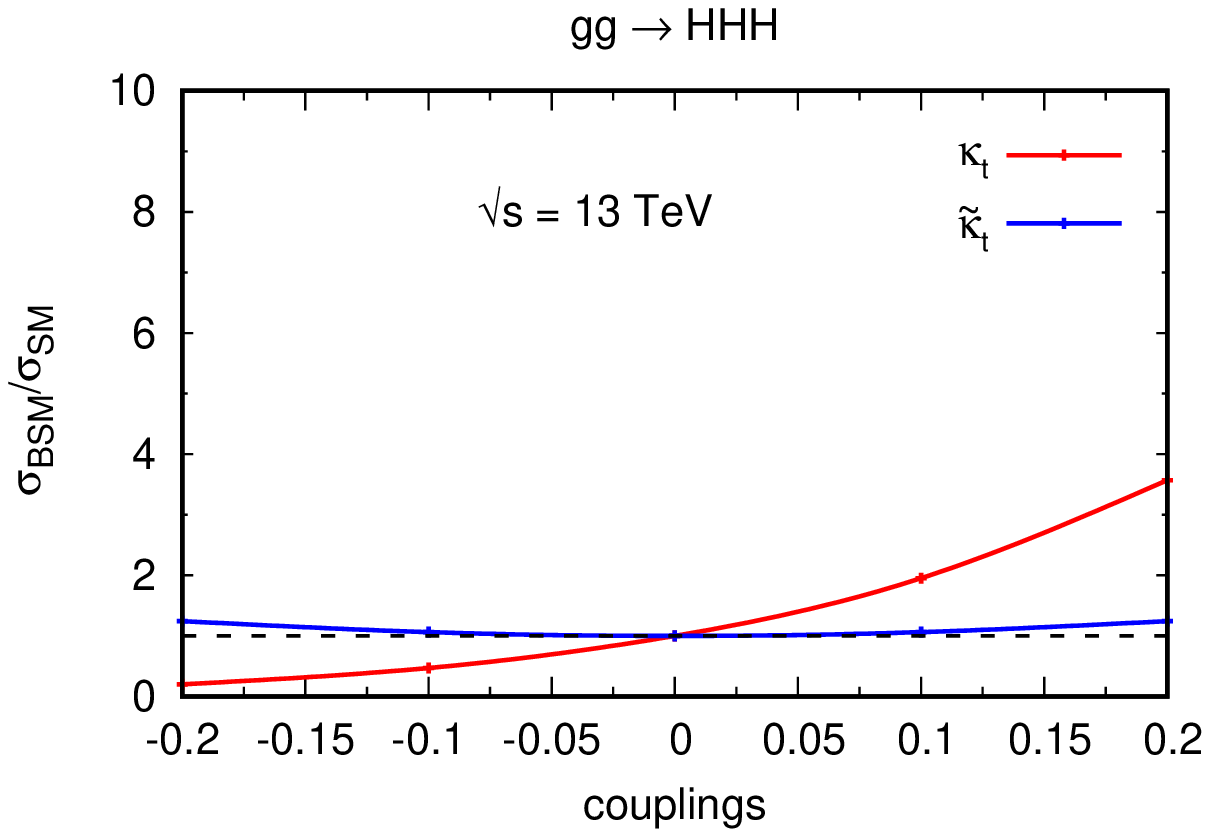}
\includegraphics [angle=0,width=0.45\linewidth]{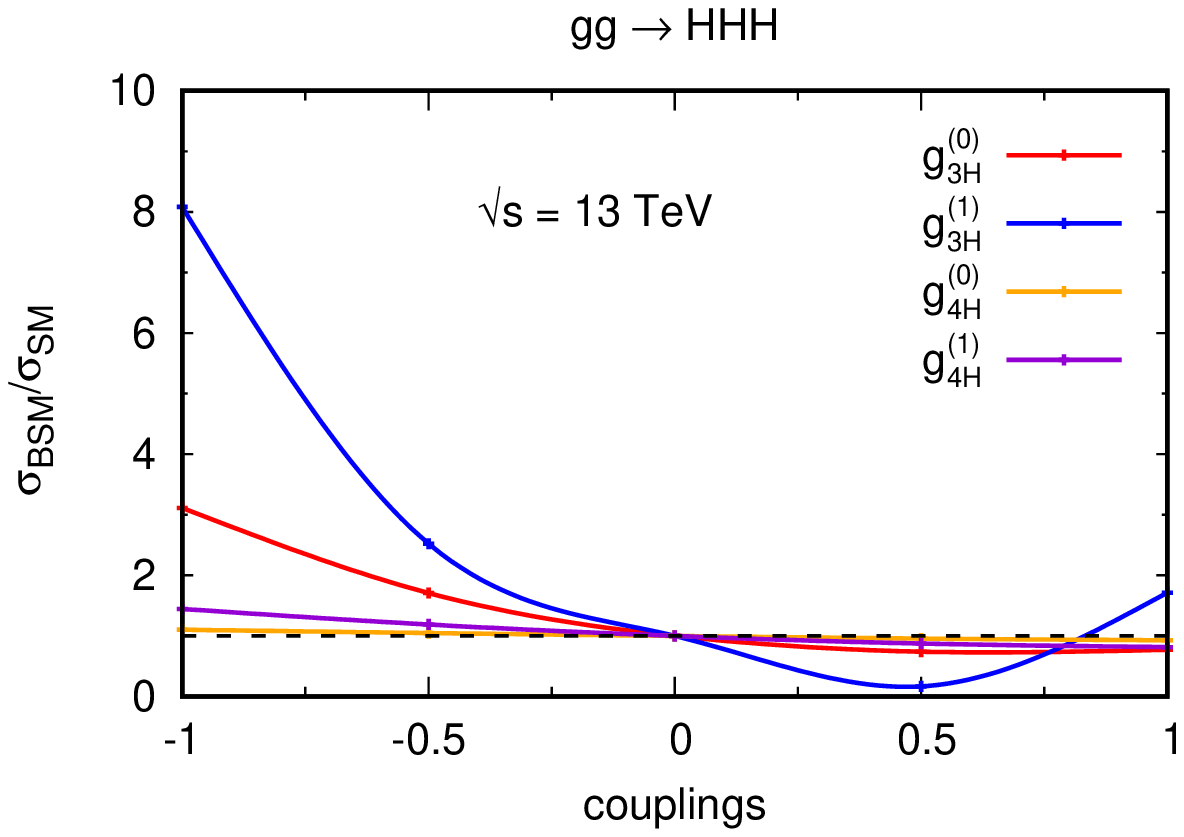}
\caption{ $\dfrac{\sigma_{\rm BSM}}{\sigma_{\rm SM}}$ as function of various Higgs anomalous 
couplings affecting $\GGHHH$ at 13 TeV.  }
\label{fig:xs-hhh-anml}
 \end{center}
\end{figure}

 \end{center}

 In the Table~\ref{table:xs-hhh-pen-bx-tr}, we have displayed the values of cross sections when only
pentagon, or box, or triangle type diagrams are considered. These categories of diagrams are separately
gauge invariant with respect to the gluons. So if we wish to estimate cross section by only keeping a category
of diagrams, we will make an order of magnitude error. This is because there is
large destructive interference among these categories of diagrams. If for simplicity,
we include only triangle diagrams in the calculation, we will underestimate the
cross section, while inclusion of only box or pentagon diagrams will overestimate
the cross section. For 13 TeV centre-of-mass energy we see that the total cross section
is about 32.0 attobarn, whereas penta, box, and triangle class  contribute
94.4, 53.6, and 3.5 attobarn respectively. This shows that there is strong destructive
interference between the different categories of diagrams. 
{At higher center-of-mass energies the interference effect becomes smaller.}
If one takes the Higgs effective field theory approach, there would be 
inclusion of triangle type diagrams only, and the cross section would be underestimated.

In Fig.~\ref{fig:dist-hhh-pen-bx-tr}, we have plotted the contribution of various category of diagrams
with respect to the $p_T$ of the leading Higgs boson at $\sqrt{s} = 13$ TeV and 100 TeV. We see that it is the pentagon
 type diagrams, which give harder Higgs boson, than other categories of diagrams.
 Interference kills such events and the $p_T$ peaks between 130 and 160 GeV.
In Fig.~\ref{fig:dists-hhh-sm}, we have plotted a number of physical quantities
 involving leading, next-to-leading, and next-to-next-to-leading Higgs bosons arranged
according to their transverse momenta. As would be expected leading Higgs boson's
$p_T$ is harder and peaks around 140-160 GeV. Softest Higgs boson $p_T$ is mostly
around 50 GeV with a large tail. However all three types are produced mainly centrally. The leading
and next-to-leading Higgs bosons are produced more back-to-back than other
Higgs bosons. Two of the softer Higgs bosons are produced closer to each other.
The masses of the Higgs boson pair also show expected behavior. The mass of the
 two larger $p_T$ Higgs bosons has a peak about 375 GeV, while for the two softer
 Higgs boson, it is near the twice of the mass of the Higgs boson. The invariant 
 mass of the three Higgs bosons peak around 550 GeV.
At higher center-of-mass energy, 100 TeV, the behavior of physical quantities
is largely the same, so we have not given separate plots.

\begin{figure}[!hbt]
\bc
\includegraphics [angle=0,width=0.45\linewidth]{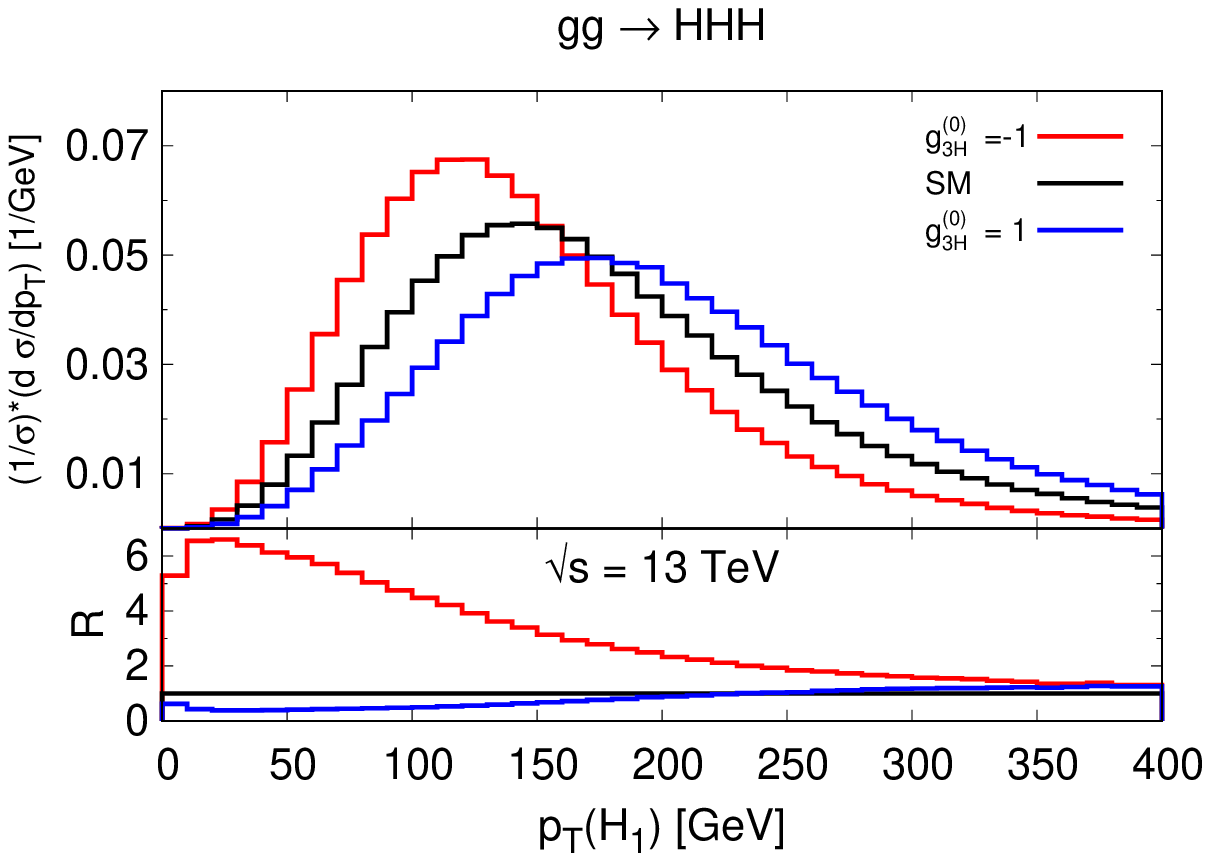}
\includegraphics [angle=0,width=0.45\linewidth]{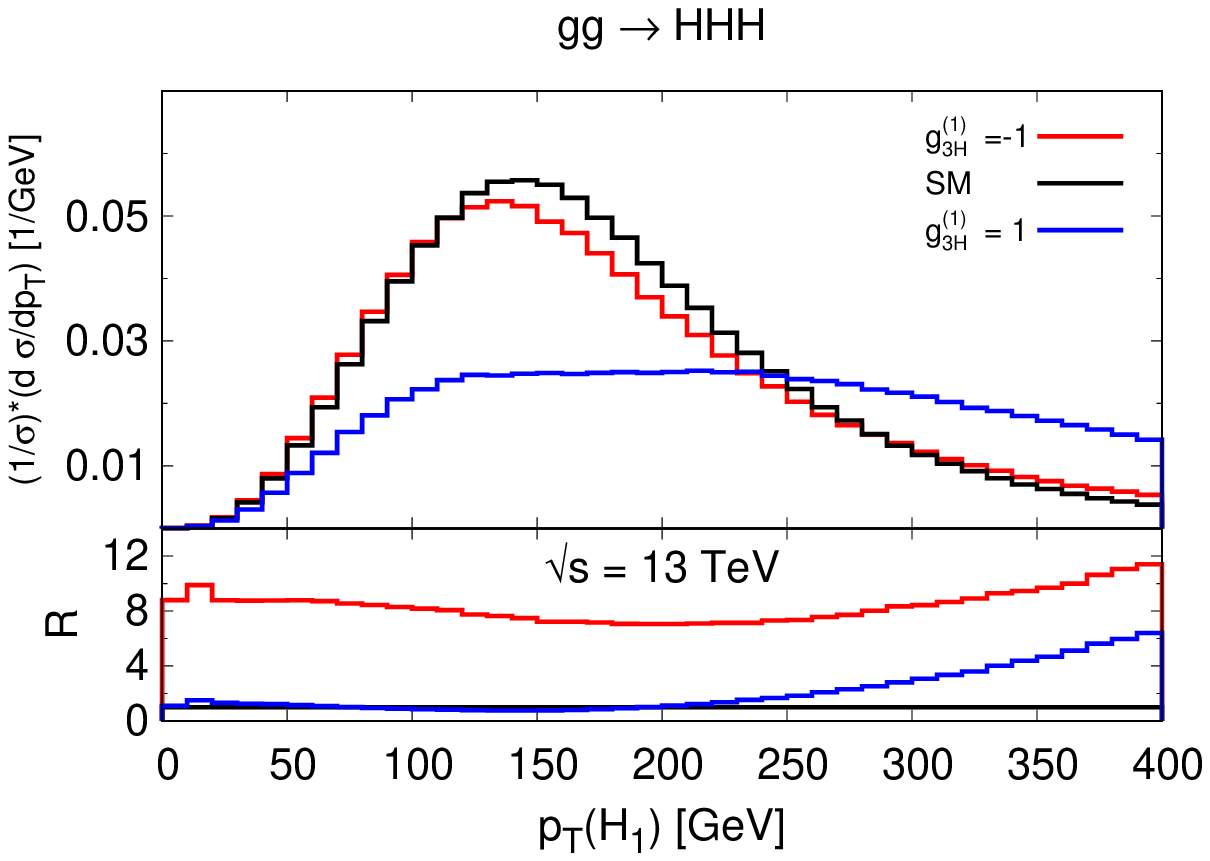}
\ec
\caption{Normalized leading $p_T(H)$ distribution in $\GGHHH$ at 13 TeV for some benchmark 
values of anomalous trilinear Higgs self-coupling. In the  
lower panels $R$ is defined as the ratio of the distributions ($d\sigma/dp_T$) in BSM and in SM. 
} \label{fig:dists-hhh-anml}
\end{figure}

In Fig.~\ref{fig:xs-hhh-anml}, we have plotted the ratios of the cross section with various anomalous
couplings and the standard model cross section. We have plotted for the range of parameters mentioned in section~\ref{sec:anml}, except for ${\tilde \kappa}_\tin{t}$ for which we have doubled the range. We see that the cross section is not sensitive to the
pseudo-scalar $ttH$ coupling and modification of quartic Higgs boson coupling.
It is not surprising because of three Higgs boson coming out of same vertex,
requiring the fourth Higgs boson to be far off-shell. 
However, it is sensitive to the scaling of the scalar $ttH$ coupling. Cross section
can significantly change with the change in this coupling -- by a factor of 3-4. The 
cross section is also sensitive to the sign of this coupling.
Interestingly, the  cross section is sensitive to trilinear Higgs boson
coupling too. Cross section can increase by an order of magnitude by the change of
derivative coupling, $g^{(1)}_{\text{\tiny 3}\tin{H}}$. This coupling is momentum dependent and will have larger effect
at higher center-of-mass energy machine. The scaling of the coupling, $g^{(0)}_{\text{\tiny 3}\tin{H}}$, can also change the 
cross section by about a factor of 3. If one could detect $HHH$ events at a future collider, one 
can probe trilinear coupling easily. Looking at Fig.~\ref{fig:dists-hhh-anml}, we see that the scaling of the coupling
changes the low $p_T$ (of the leading Higgs boson) events more; while derivative coupling tends to also change
large $p_T$ events. Therefore, one can probe both couplings by focusing on low
$p_T$ events in one case and higher $p_T$ events in another case. The $p_T$ distributions for
the sub-leading Higgs boson and mass distributions of two or three Higgs bosons show similar
sensitivity. Note that the cross-section is symmetric for ${\tilde \kappa}_\tin{t}$.

Let us now consider the possibility of observing the production of 
the three Higgs bosons, $HHH$. At the LHC,
the cross section is of the order of $32$ ab.  It leads to too few events with even
the highest possible integrated luminosity of $3$ ab$^{-1}$ to be observable in the
sea of the background. However we have seen that this process is sensitive to
trilinear Higgs boson self-coupling. Such anomalous couplings can enhance the cross section by a
factor of 3 to 8. Even with this enhancement, once we include branching ratios,
kinematic cuts, tagging, and other efficiencies, there will be too few events to
be visible. Given that it is one of the few processes to help determine the
quartic Higgs boson self-coupling, it will be worthwhile to look for this process.
At a 100 TeV machine, it could be possible. The cross section there is about
$3100$ ab. With large enough data, this process might be observable. The process
$pp \to HHH$ will give rise to `multileptons' and `few leptons with jets' signatures.
There will be irreducible background from $ZZZ$ production, and an array of
reducible backgrounds depending on the signature. The $ZZZ$ production cross
section is about $136$ fb which can be controlled if we include branching
ratios and construct Higgs boson masses. In the case of reducible backgrounds
there will be background from a top-pair production with jets or vector bosons, multi
vector boson production with jets, and multijets. To tame these backgrounds,
one may require tagging of bottom jets and tau jets. Given the severity of the
background, one may still need a multivariate analysis. In \cite{Papaefstathiou:2015paa},
authors have studies the the channel $HHH \to b {\bar b} b {\bar b} \gamma \gamma$ in the standard
model and its simple deformation with marginal success.  More recently the authors of
\cite{Fuks:2017zkg} have studied `$b {\bar b} b {\bar b} \tau \tau$' channel
and authors in \cite{Kilian:2017nio} have similarly studied `$b {\bar b} l^{+} l^{-} + 4 \;{\rm jets}$'
channel with modest success. Therefore, a modification
of the interactions that will enhance the signal significantly and improved search
strategies will be needed to determine Higgs boson self-couplings.

%
%
\subsection{The process $pp \to HHZ$}\label{sec:results-hhz}

   Unlike the $p p \to HHH$ process, this process can occur at the tree level. In this section, we
   will mainly focus the NNLO contribution to this process that occur via $g g \to HHZ$ process. 
   We have estimated tree level value and one-loop QCD corrections, i. e., NLO contribution
   using {\tt MadGraph5\_aMC@NLO} \cite{Alwall:2014hca}. We will see that NNLO contribution is comparable to NLO contribution at 13 TeV. 
   NNLO contribution becomes even more important as the center-of-mass energy increases, and can 
   become comparable to the tree level value.

\begin{table}[!hbt]
 \begin{center}
  \begin{tabular}{|c|c|c|c|c|c}
   \hline
   $\sqrt{\rm s}\;(\rm TeV)$& $8$& $13$& $33$& $100$ \\
   \hline
   \hline
   & & & &\\
     $\sigma^{\tin{ \rm HHZ,\,LO}}_{\tin{\rm  GG}}\;~[\rm ab]$
     & $10.0^{+34.0\%}_{-24.0\%}$ & $42.3^{+30.9\%}_{-21.4\%}$ & $406.7^{+23.9\%}_{-17.9\%}$ & $3562.4^{+16.8\%}_{-13.9\%}$ \\
   & & & &\\
    $\sigma^{\tin{\rm HHZ,\,LO}}_{\tin{\rm QQ}}\;~[\rm ab]$
    & $97.2^{+3.9\%}_{-3.8\%}$ & $236.7^{+1.3\%}_{-1.5\%}$ & $988.8^{+2.6\%}_{-3.3\%}$ & $4393.0^{+7.1\%}_{-7.8\%}$ \\
   & & & &\\
     $\sigma^{\tin{\rm HHZ,\,NLO}}_{\tin{\rm QQ}}\;[\rm ab]$
     & $122.0^{+1.7\%}_{-1.6\%}$  & $ 294.5^{+1.5\%}_{-1.0\%}$ & $1197.0^{+1.7\%}_{-1.9\%}$ & $4971.0^{+1.8\%}_{-3.2\%}$ \\
   & & & & \\
   \hline \hline
    & & & & \\
     ${\rm R}_{1} = \frac{\sigma^{\sbx{.4}{\textit{\rm HHZ,\,LO}}}_{\sbx{.4}{\rm GG}}}{\sigma^{\sbx{.4}{\textit{\rm HHZ,\,LO}}}_{\sbx{.4}{\rm QQ}}}$
     & 0.10 & 0.18 & 0.41 & 0.81  \\
         & & & & \\
     ${\rm R}_{2} = \frac{\sigma^{\sbx{.4}{\textit{\rm HHZ,\,LO}}}_{\sbx{.4}{\rm GG}}}{\sigma^{\sbx{.4}{\textit{\rm HHZ,\,NLO}}}_{\sbx{.4}{\rm QQ}}}$
     & 0.08 & 0.14 & 0.34 & 0.72  \\
   & & & & \\
     ${\rm R}_{3} = \frac{\sigma^{\sbx{.4}{\textit{\rm HHZ,\,LO}}}_{\sbx{.4}{\textit{\rm GG}}}}{({\sigma}^{\sbx{.4}{\textit{\rm HHZ,\,NLO}}}_{\sbx{.4}{\textit{\rm QQ}}}-\sigma^{\sbx{.4}{\textit{\rm HHZ,\,LO}}}_{\sbx{.4}{\textit{\rm QQ}}})}$
     & 0.40 & 0.73 & 1.95 & 6.16  \\
   & & & &\\
   \hline
  \end{tabular}
 \end{center}
 \caption{ A comparison of different perturbative orders in QCD coupling contributing to $pp \to HHZ$
hadronic cross section at $\sqrt{s}=8, 13, 33$, and 100 TeV. We also calculate ratios 
$R_1$, $R_2$ and $R_3$ which quantify the GG contribution with respect to the $QQ ({\rm LO})$ 
and $QQ({\rm NLO})$ contributions. 
} \label{table:xs-hhz}
\end{table}

  In Table~\ref{table:xs-hhz}, we have given LO, NLO, and NNLO contribution to this process at different center-of-mass
    energies. We have used {\tt cteq6l1} parton distribution for LO and NNLO calculation and {\tt cteq6m} for the
    NLO calculation \cite{Nadolsky:2008zw}, with $\hat{s}$ as renormalization/factorization scales for NNLO calculation and sum of transverse
    mass for LO and NLO calculation (in {\tt MadGraph5\_aMC@NLO}). Uncertainties are
    estimated by using other cteq6 parton distributions and changing the renormalization/factorization 
    scales by a factor of 2. 
    At the LHC, the leading order contribution is about 237 ab. 
    NLO corrections are about $24\%$, and add 57.8 ab to the LO value. NNLO corrections are about 
    $18\%$, and add 42.3 ab to the LO value. The cross section including LO, NLO, and NNLO values is
    about 336.8 ab, leading to about 1000 events at the maximal proposed integrated luminosity. As in the
    case of $HHH$ production, the value of NNLO contribution becomes more significant as center-of-mass
    energy increases. At 100 TeV, NLO correction is only $13\%$, while NNLO contribution is 
    $81\%$ of the LO value due to an increase in the gluon flux. We see that NNLO contribution approaches the LO value. It is however 
    not alarming, only more useful. NNLO contribution is from gluon-gluon annihilation, 
    while LO result is from quark-antiquark scattering. At 13 TeV, the scale uncertainties 
    for GG channel are in the range of -21\% to 31\%. { The reason for this large uncertainty and its decrease with center of mass energy is same as that for $gg \to HHH$ process. Uncertainties in the QQ processes are smaller, as these are primarily electroweak processes.}

\begin{table}[!hbt]
 \begin{center}
  \begin{tabular}{|c|c||c||c||c|}
   \hline
   $\sqrt{\rm s}\;(\rm TeV)$ & 8 & 13 & 33 & 100 \\[3pt]
   \hline
   & & & &\\
   $\sigma^{\rm \tiny HHZ}_{\rm \tiny penta} \; [\rm ab]$ & 30.8 & 148.1 & 1718.4 & 17694.0   \\
& & & &\\
   \hline
   & & & &\\
   $\sigma^{\rm \tiny HHZ}_{\rm \tiny box} \; [\rm ab]$ & 73.1 & 434.7 & 7468.2 & 115747.2  \\
& & & &\\
   \hline
& & & &\\
   $\sigma^{\rm \tiny HHZ}_{\rm \tiny triangle} \; [\rm ab]$ & 78.4 & 475.6 & 8157.2 & 124273.1 \\
& & & &\\
   \hline \hline
   & & & &\\
   $\sigma^{\rm \tiny HHZ}_{\rm \tiny total} \; [\rm ab]$ & 10.0 & 42.3 & 406.4 & 3557.5 \\
& & & &\\
   \hline
  \end{tabular}
 \end{center}
  \caption{SM contribution of pentagon, box, and triangle diagrams to the total cross section in $\GGHHZ$ 
  at different collider center-of-mass energies, displaying a destructive interference effect. }
  \label{table:xs-hhz-pen-bx-tr}
\end{table}

\begin{figure}[!hbt]
\bc
\includegraphics [angle=0,width=0.45\linewidth]{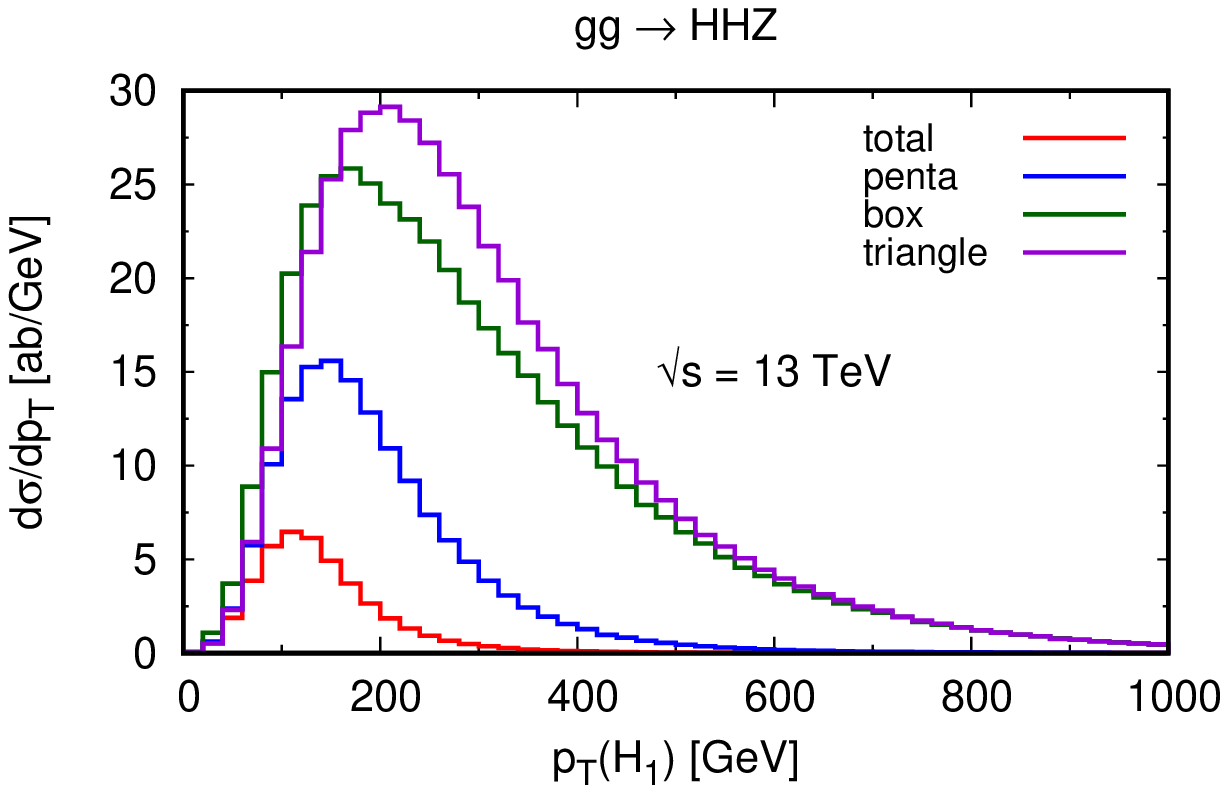}
\includegraphics [angle=0,width=0.45\linewidth]{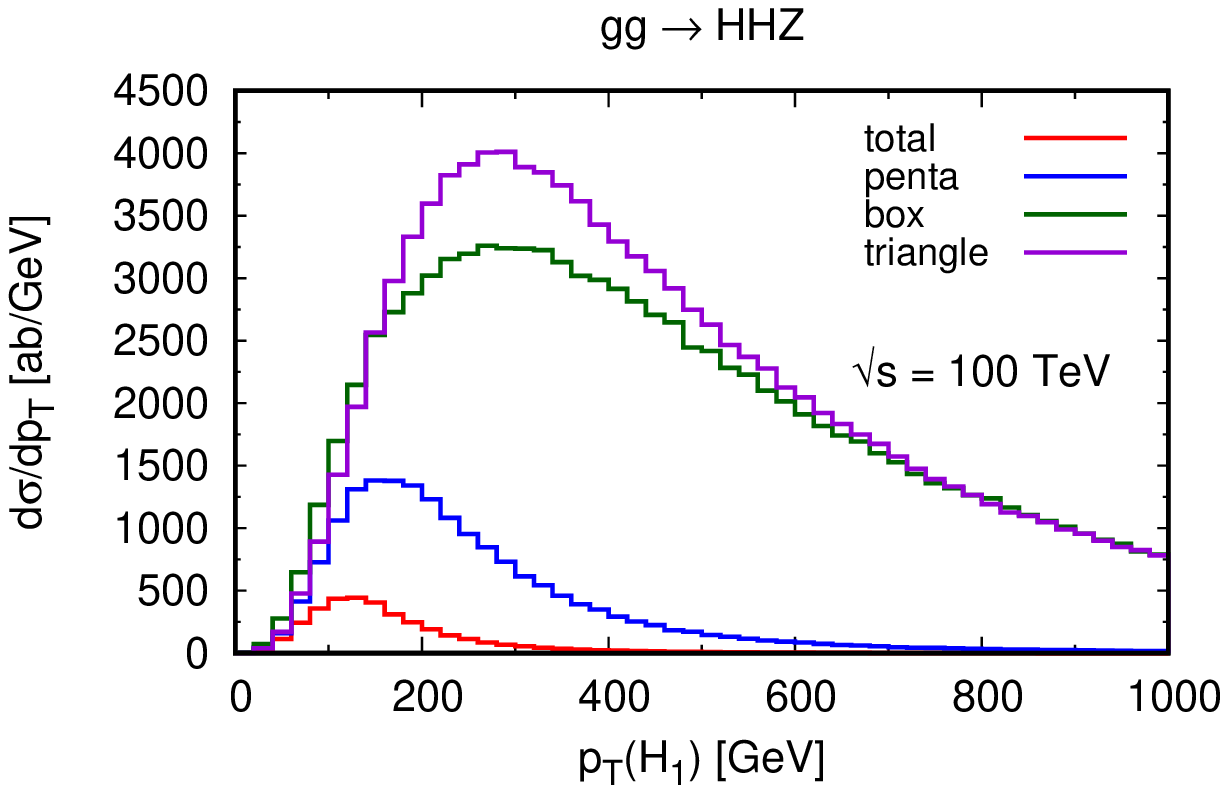}
\ec
\caption{ SM contribution of pentagon (blue), box (green) and triangle (violet) diagrams to 
leading $p_T(H)$ distribution in $\GGHHZ$ at 13 TeV (left) and 100 TeV (right).} \label{fig:dist-hhz-pen-bx-tr}
\end{figure}

\begin{figure}[!hbt]
\bc
\includegraphics [angle=0,width=0.45\linewidth]{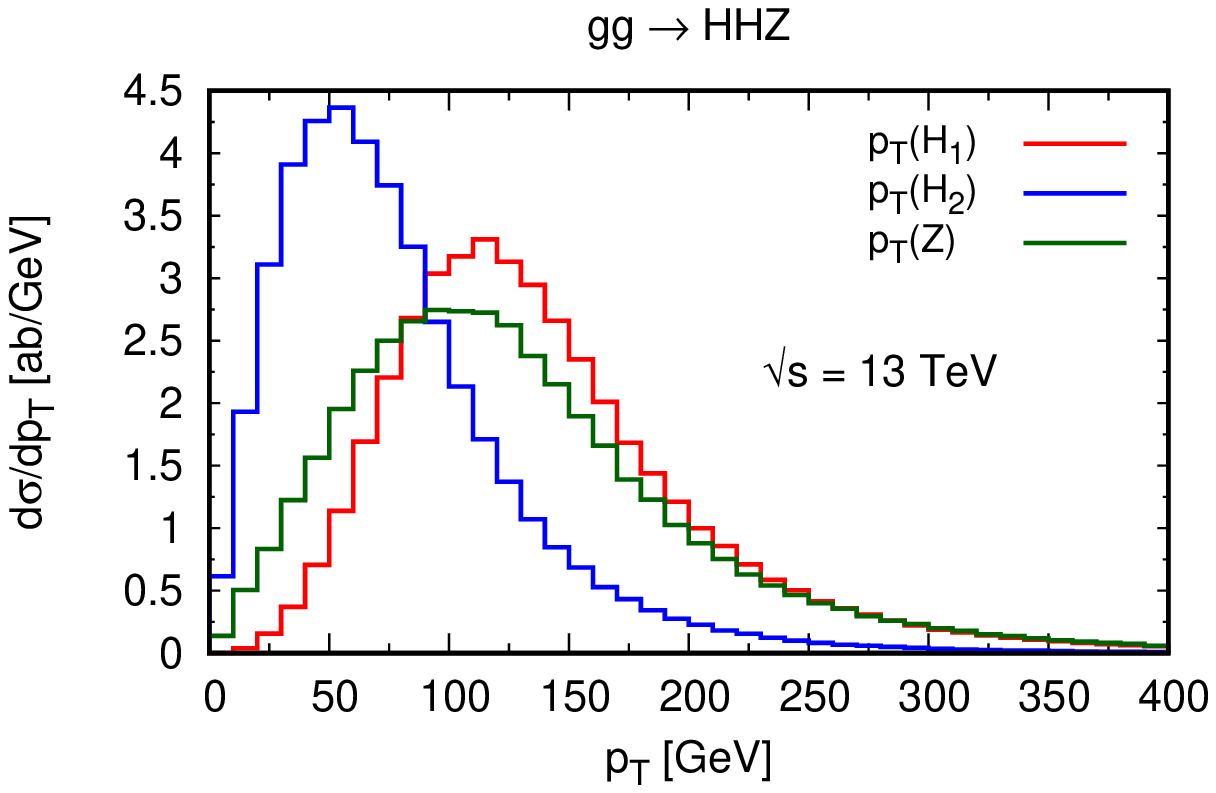}
\includegraphics [angle=0,width=0.45\linewidth]{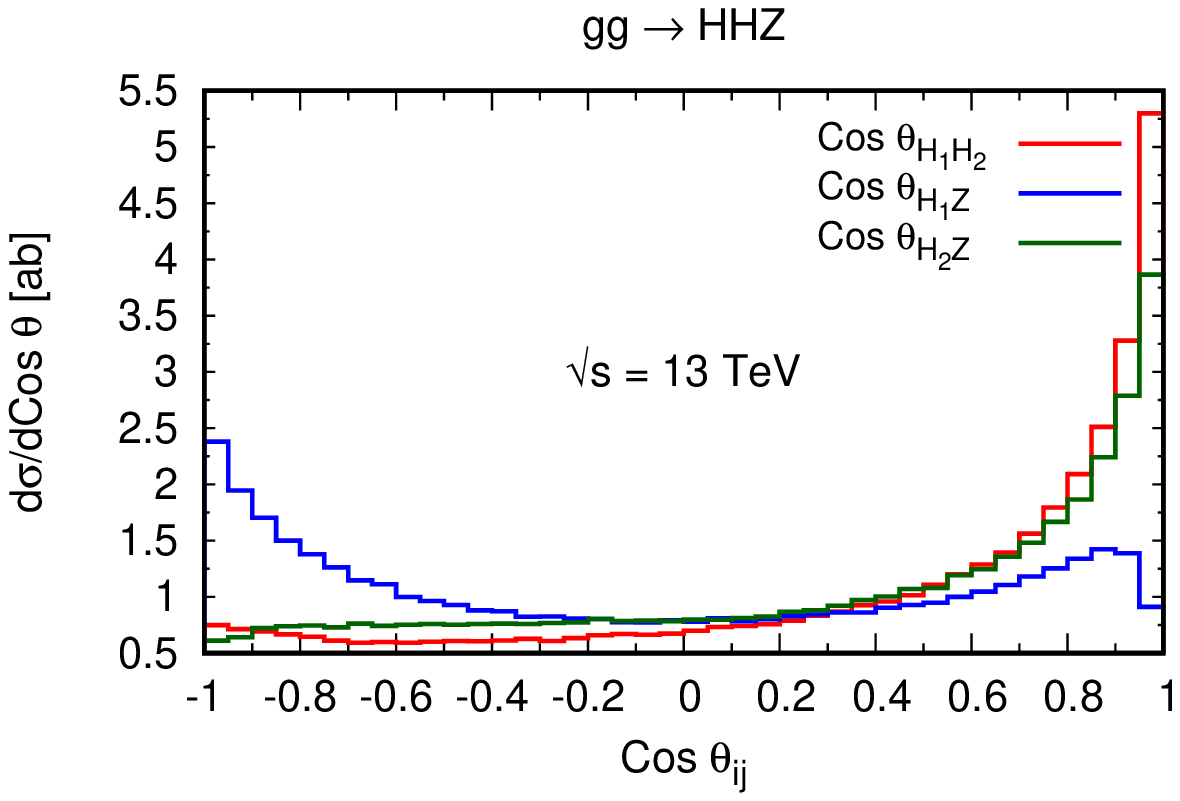}
\includegraphics [angle=0,width=0.45\linewidth]{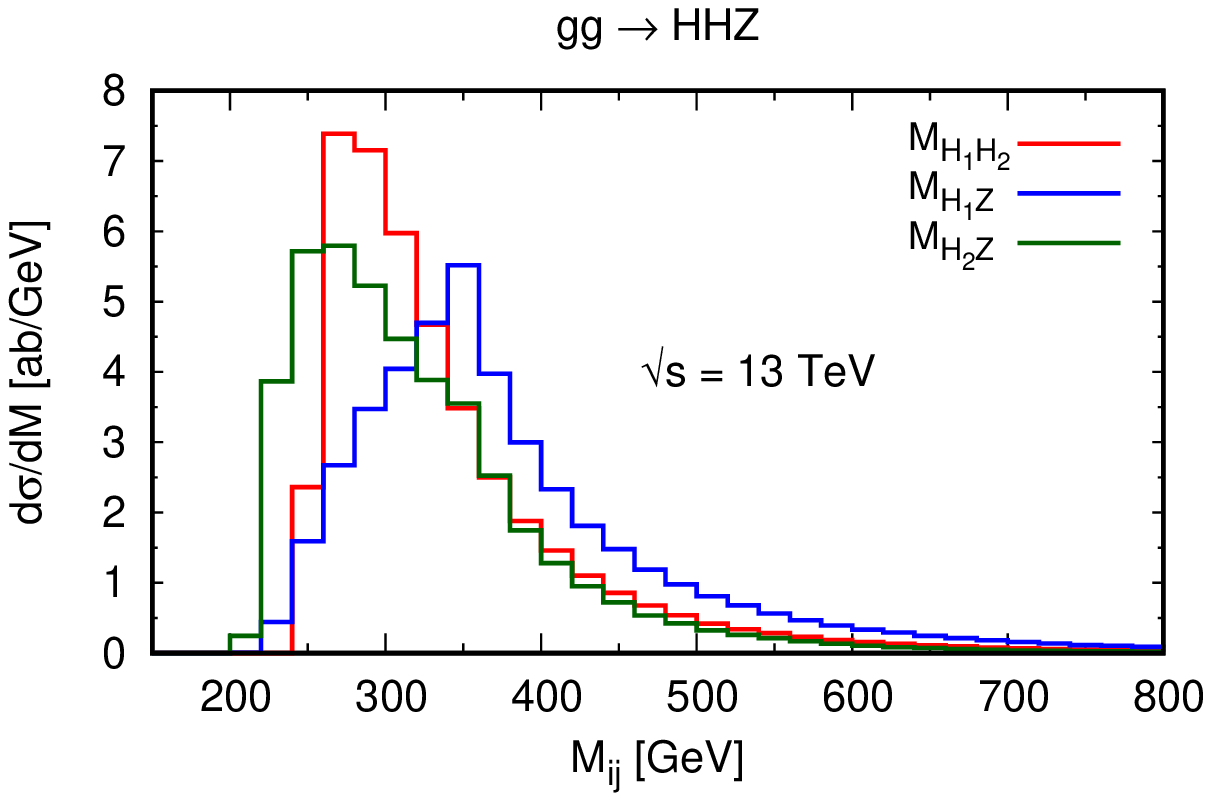}
\includegraphics [angle=0,width=0.45\linewidth]{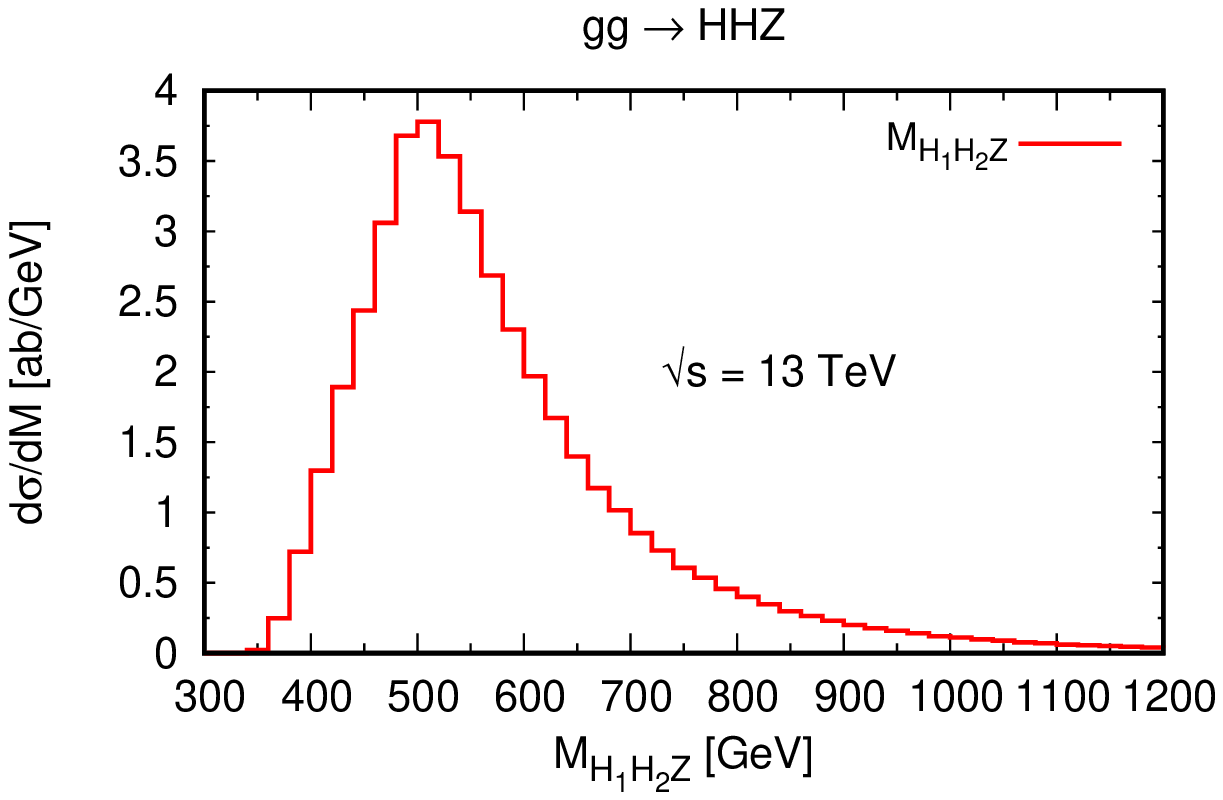}
\ec
\caption{ Kinematic distributions for $\GGHHZ$ in the SM at 13 TeV. These plots are obtained 
after $p_T$ ordering the Higgs bosons. $H_1$ and $H_2$ refer to the hardest and 
second hardest Higgs bosons in $p_T$ respectively.} \label{fig:dists-hhz-sm}
\end{figure}

\begin{figure}[!hbt]
\bc
\includegraphics [angle=0,width=0.45\linewidth]{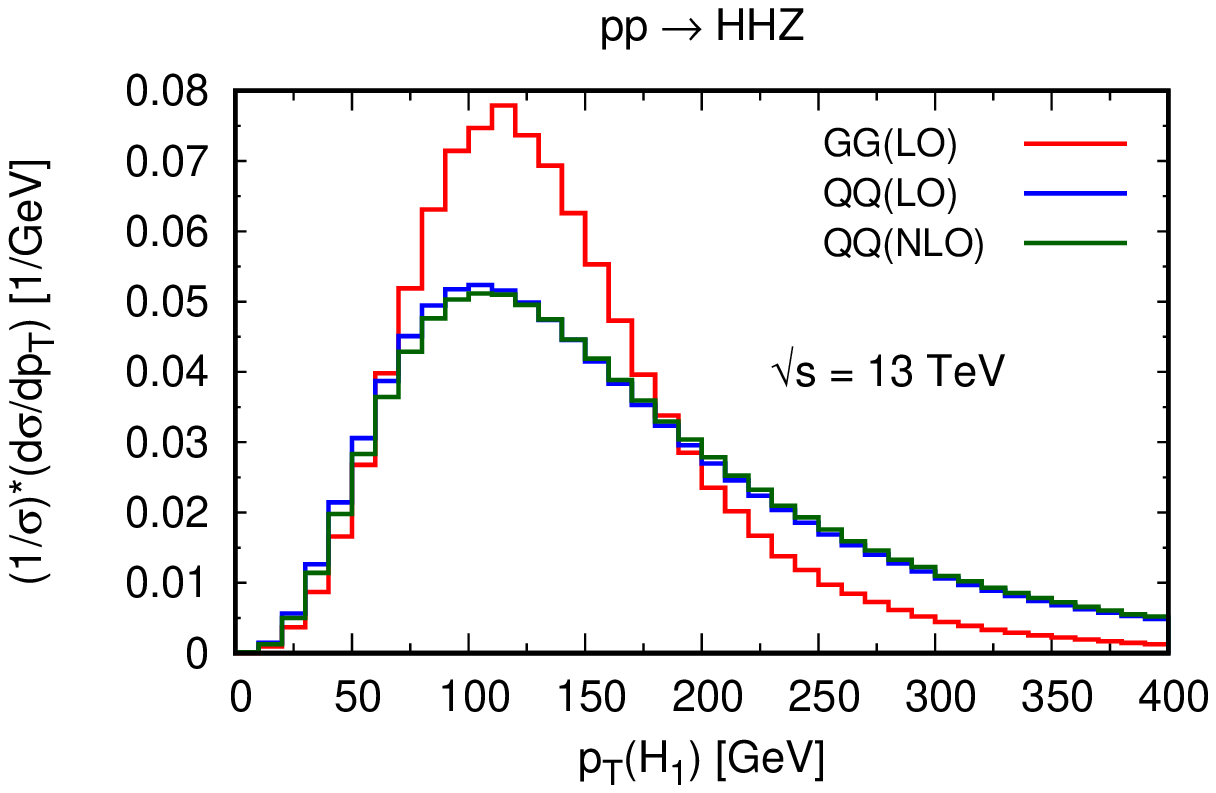}
\includegraphics [angle=0,width=0.45\linewidth]{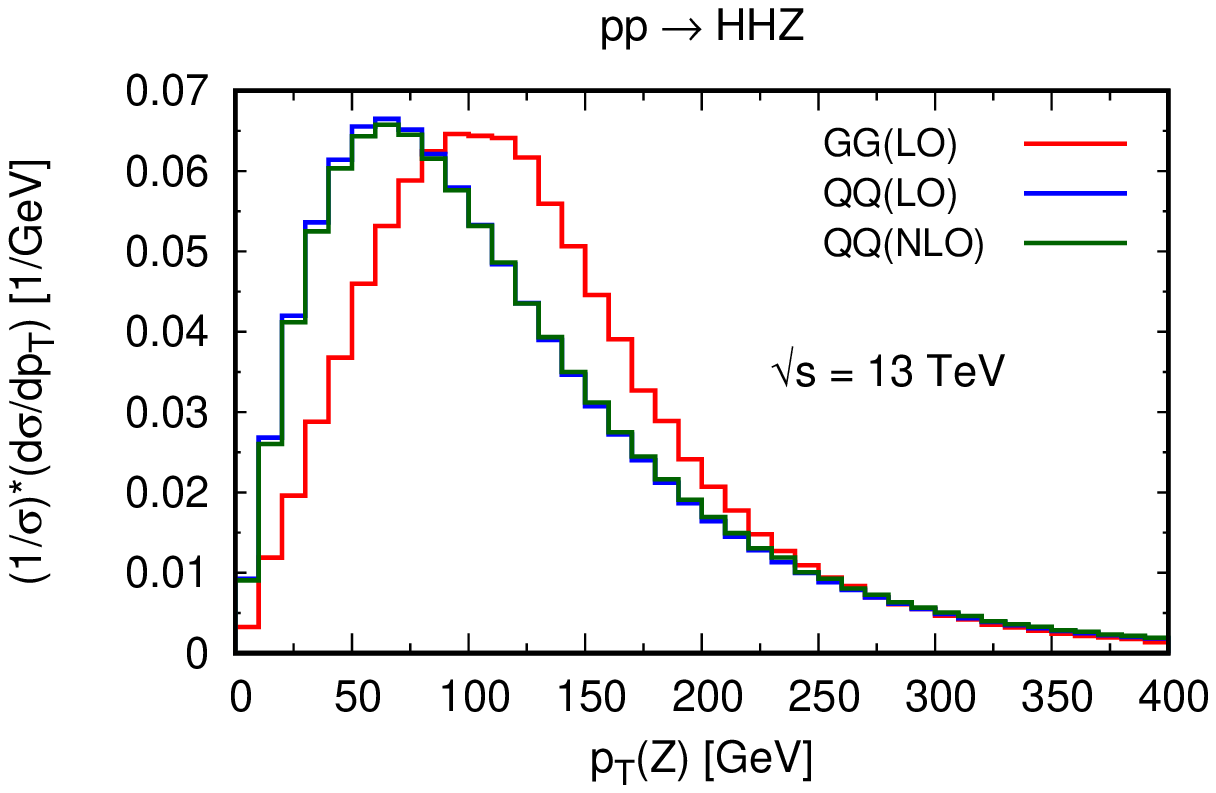}
\ec
\caption{ A comparison of normalized distributions for $p_T(H_1)$ and $p_T(Z)$ due to $\GGHHZ$ and $\QQHHZ$ 
in the SM at 13 TeV. } \label{fig:dists-qqhhz-gghhz-norm-sm}
\end{figure}    
    
\begin{figure}[!hbt]
\bc
\includegraphics [angle=0,width=0.45\linewidth]{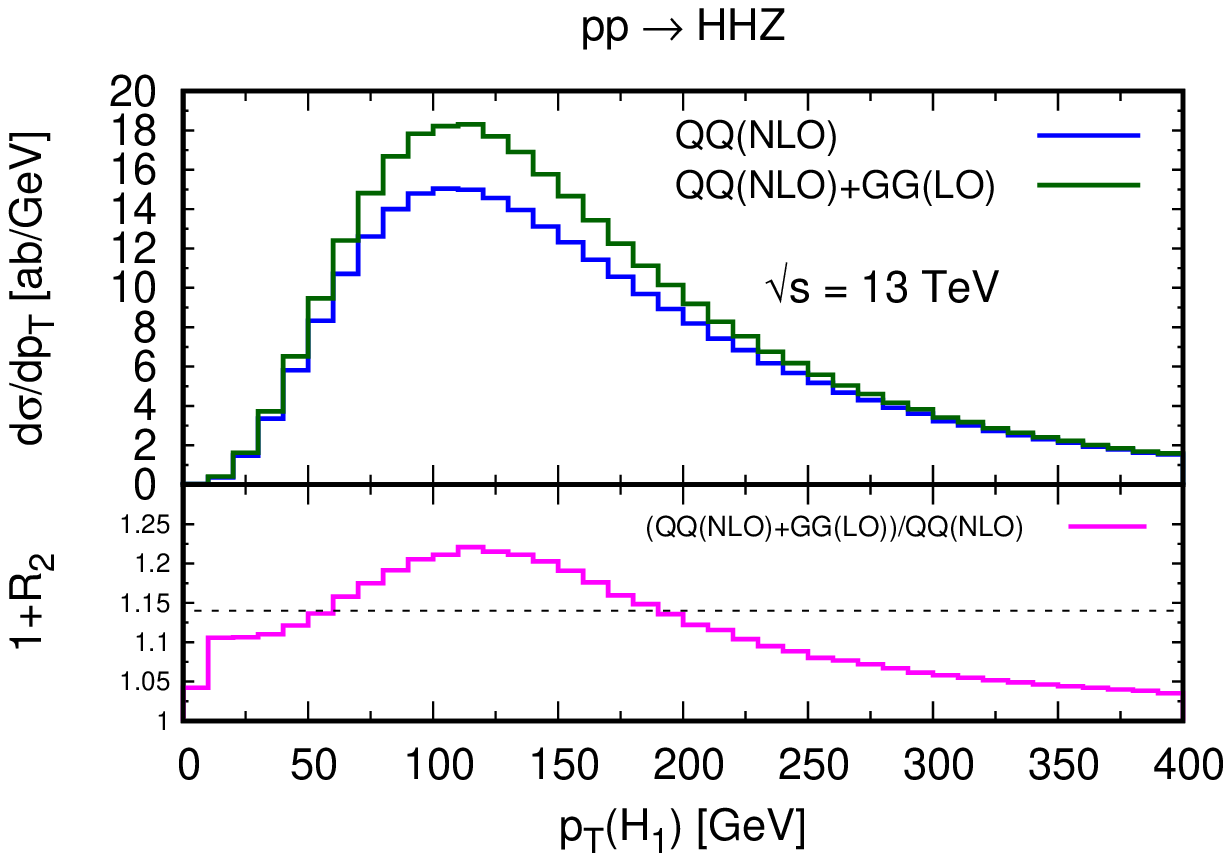}
\includegraphics [angle=0,width=0.45\linewidth]{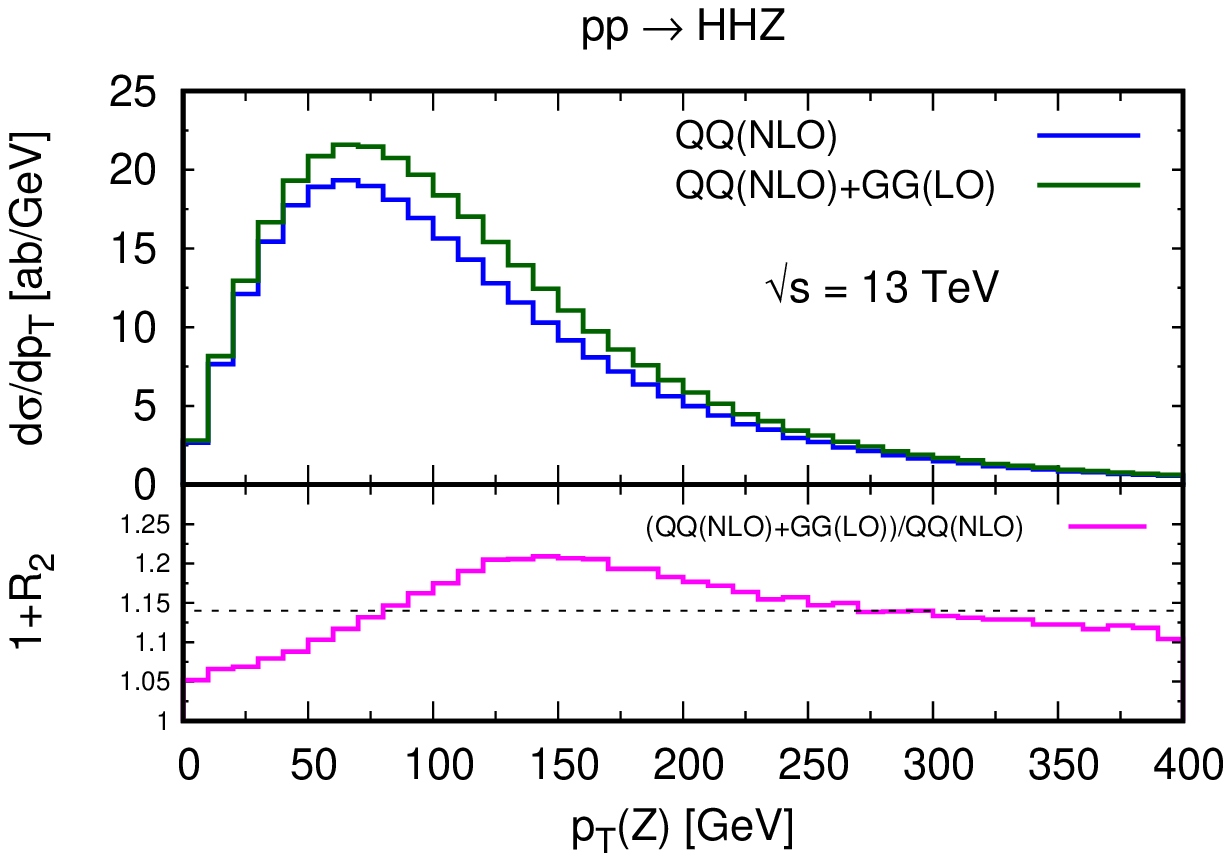} \\
\includegraphics [angle=0,width=0.45\linewidth]{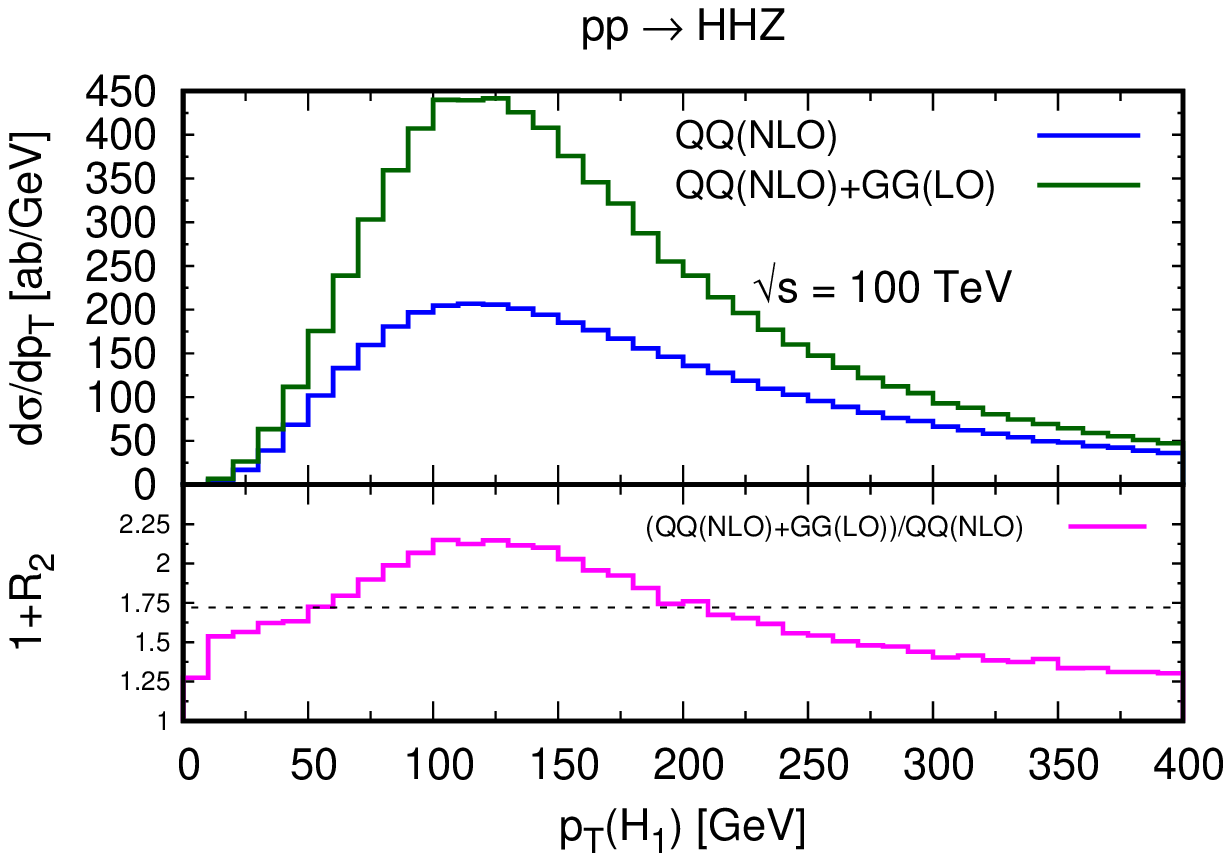}
\includegraphics [angle=0,width=0.45\linewidth]{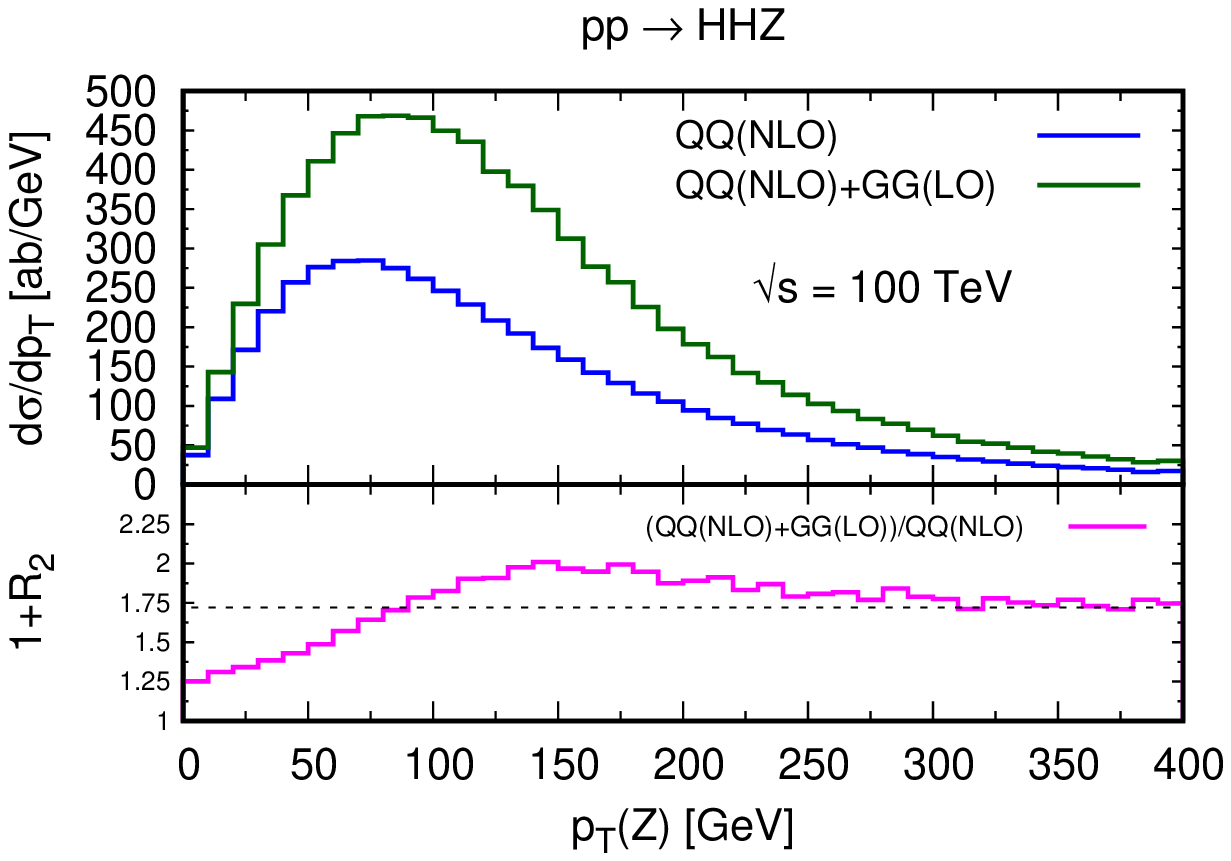}
\ec
\caption{ Combined $\GGHHZ$(LO) $+$ $\QQHHZ$(NLO) contribution to $p_T(H_1)$ and $p_T(Z)$ distributions in the SM at 13 TeV and 100 TeV. 
Lower panels show the ratio of $QQ({\rm NLO})+GG({\rm LO})$ and $QQ({\rm NLO})$ for each of these distributions. 
The dashed straight line in the lower panel
of each plot refers to the same quantity at inclusive or total cross section level (see $R_2$ in Table~\ref{table:xs-hhz}).} \label{fig:dists-qqhhz-gghhz-sm-13TeV}
\end{figure}    
  
%

\begin{figure}[!hbt]
\bc
\includegraphics [angle=0,width=0.30\linewidth]{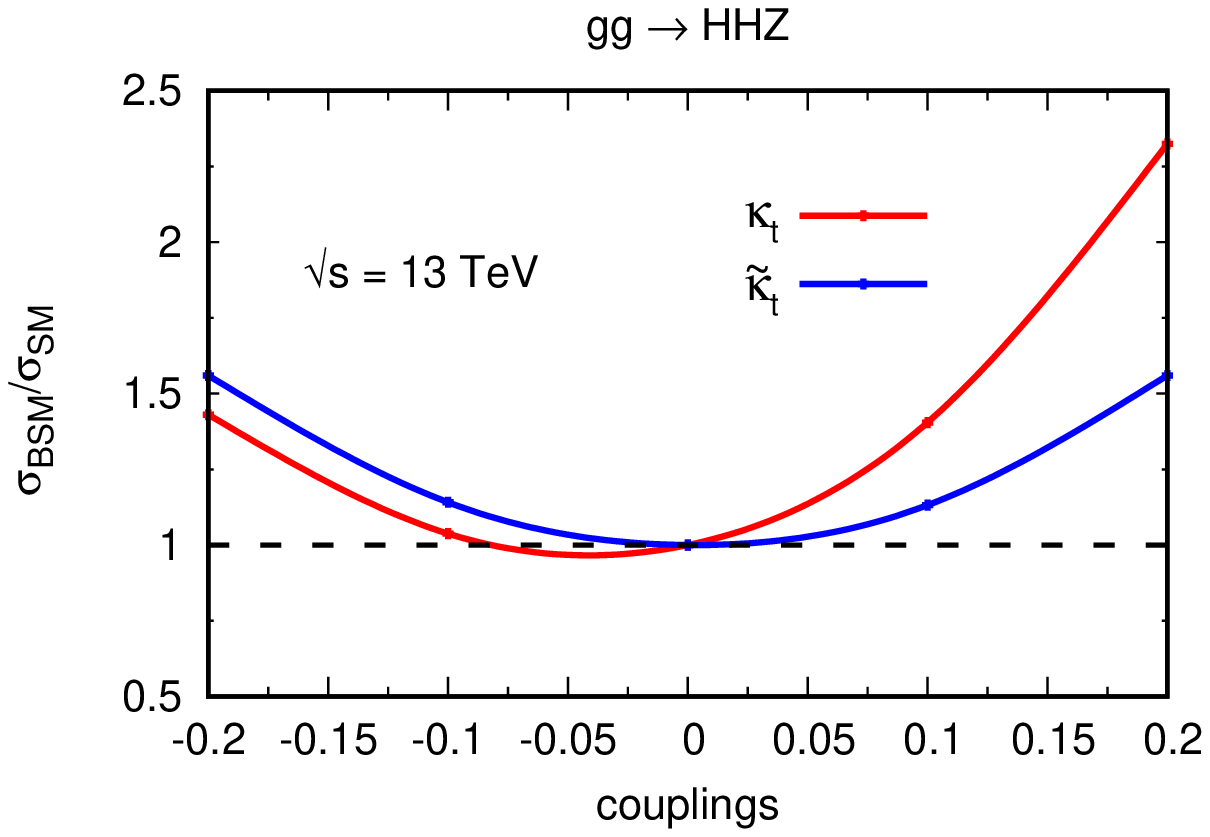}
\includegraphics [angle=0,width=0.30\linewidth]{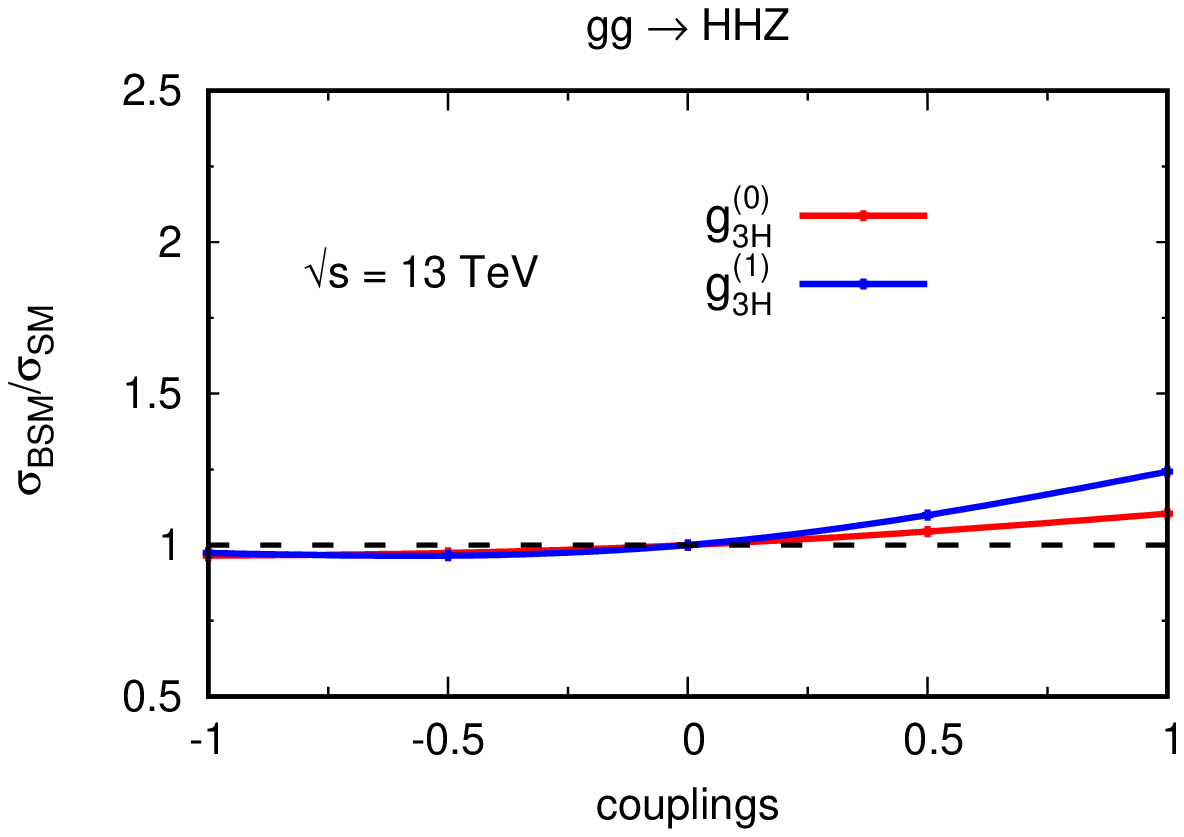}
\includegraphics [angle=0,width=0.30\linewidth]{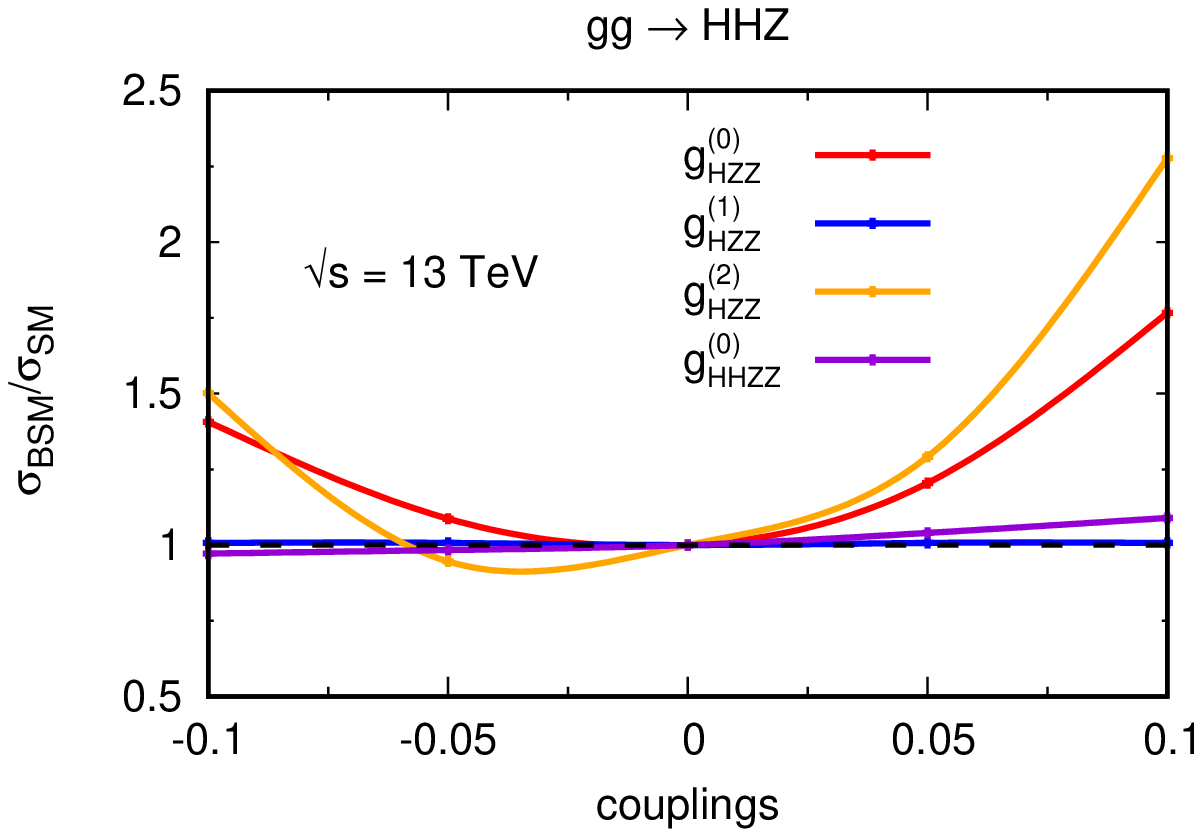}
\ec
\caption{ $\dfrac{\sigma_{\rm BSM}}{\sigma_{\rm SM}}$ as a function of anomalous couplings of the Higgs boson 
in $\GGHHZ$ at 13 TeV.  }
\label{fig:xs-hhz-anml}
\end{figure}

  Table~\ref{table:xs-hhz-pen-bx-tr} demonstrates interference effect in the process $HHZ$. Unlike $HHH$ process, here triangle
    diagram contribution is larger and interference effect is more severe. If we keep only one
    class of diagrams which are gauge invariant with respect to the gluons, we will overestimate the cross section by
     an order of magnitude at 13 TeV.  We see that the total cross section
    is 42.3 attobarn, whereas penta, box, and triangle class  contribute 148.1, 434.7,
    and 475.6 attobarn respectively. This shows there is a strong destructive interference
    between the different class of diagrams. The destructive interference effect becomes stronger at higher $\sqrt{s}$.
    In Fig.~\ref{fig:dist-hhz-pen-bx-tr}, we have plotted
    the contribution of the individual class of diagrams with respect to the $p_T$ of the 
    leading (in $p_T$) Higgs boson. Here effects are different. Triangle and box diagrams
    have larger cross section and contribute more to higher $p_T$ events. The interference effects
    seem to cut off large $p_T$ contribution of the individual class of diagrams. The $p_T$ of the
    leading Higgs boson, after the interference, shifts to lower values and peaks around 120 GeV.

\begin{figure}[!hbt]
\bc
\includegraphics [angle=0,width=0.45\linewidth]{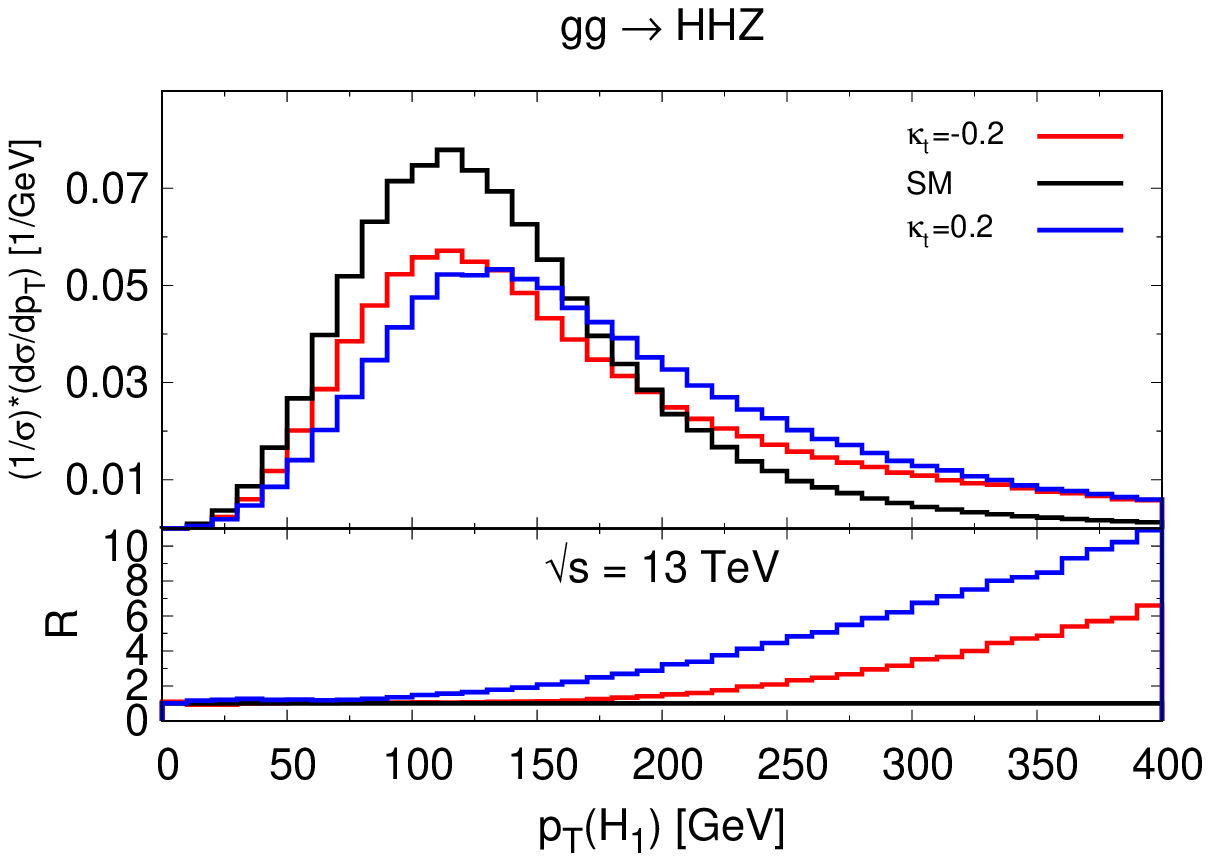}
\includegraphics [angle=0,width=0.45\linewidth]{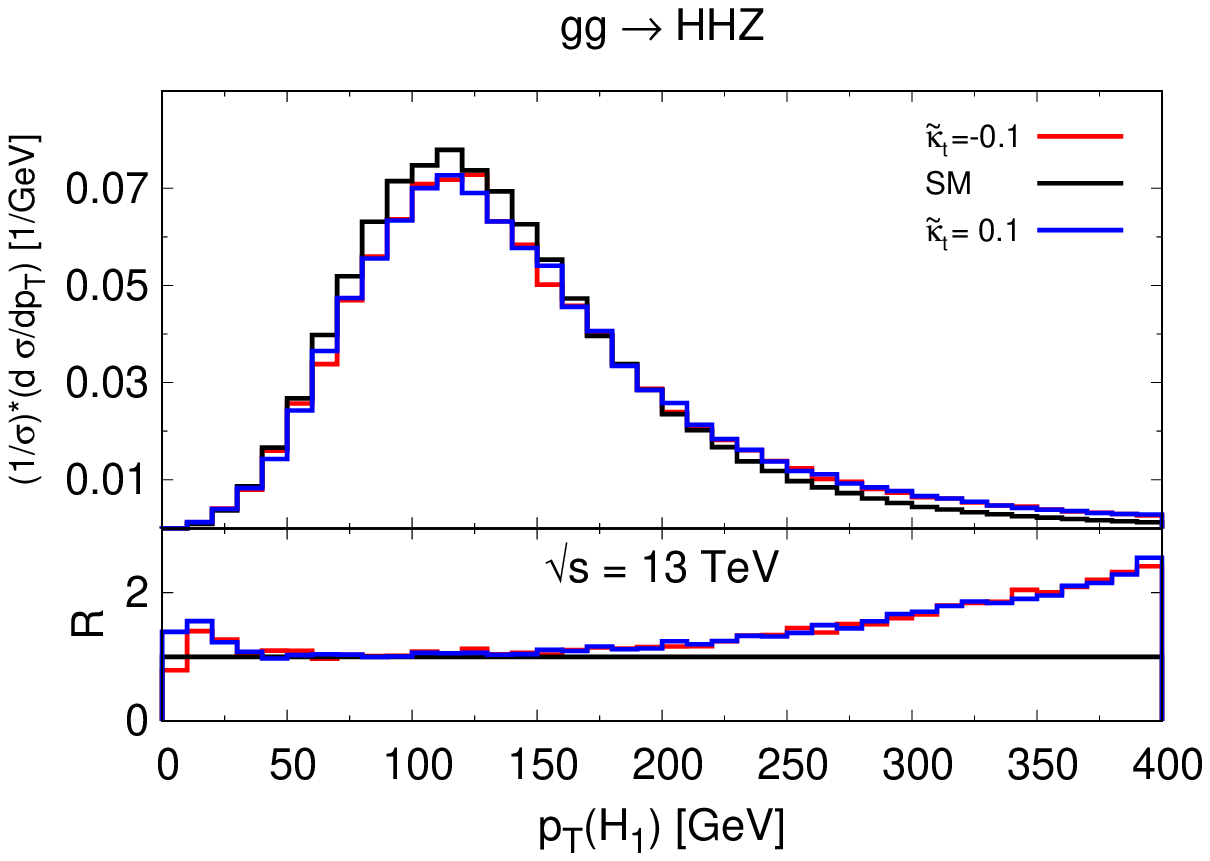}
\ec
\caption{ Normalized leading $p_T(H)$ distribution in $\GGHHZ$ at 13 TeV for some benchmark 
values of anomalous top Yukawa couplings. } \label{fig:dists-hhz-anml-yt}
\end{figure}

\begin{figure}[!hbt]
\bc
\includegraphics [angle=0,width=0.45\linewidth]{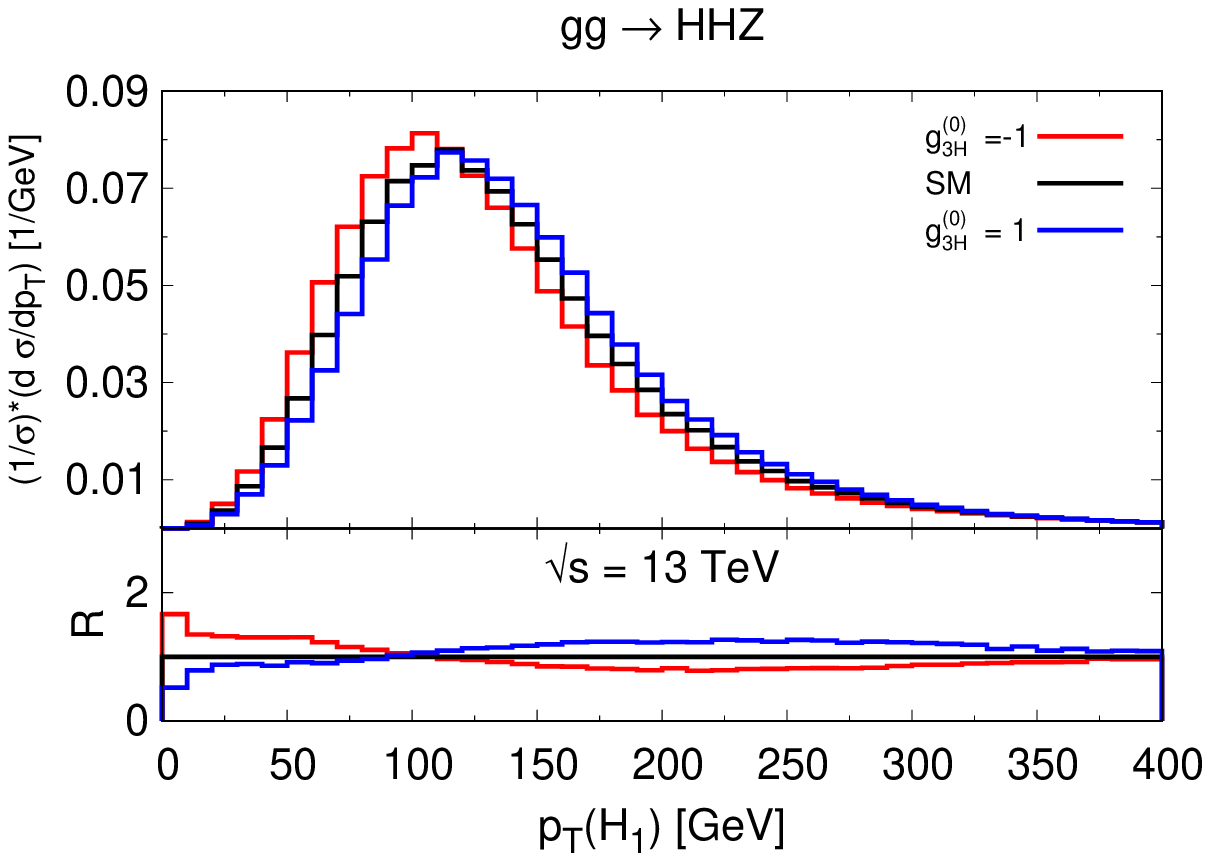}
\includegraphics [angle=0,width=0.45\linewidth]{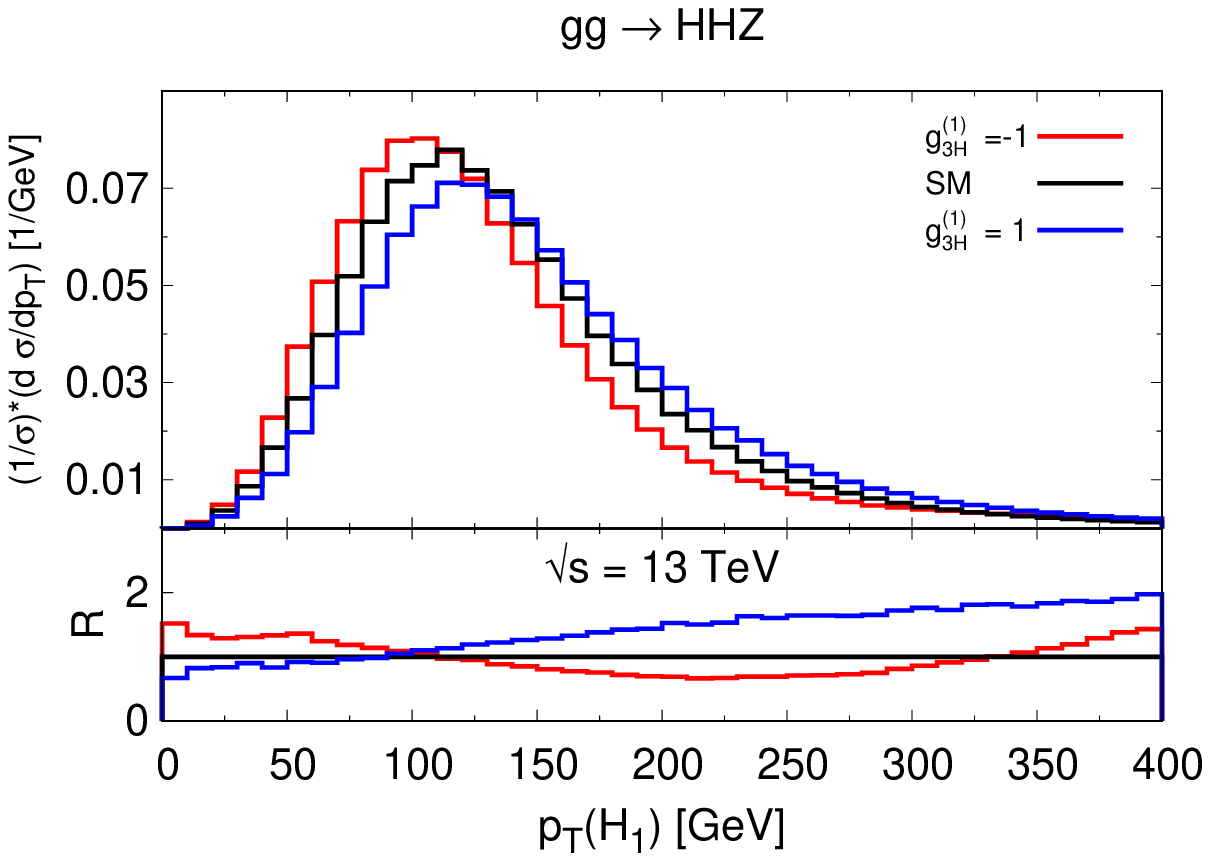}
\ec
\caption{Normalized leading $p_T(H)$ distribution in $\GGHHZ$ at 13 TeV for some benchmark 
values of $HHH$ anomalous couplings.} \label{fig:dists-hhz-anml-h3}
\end{figure}

In Fig.~\ref{fig:dists-hhz-sm}, we have plotted a few physical quantities, as in the case of $HHH$
production for 13 TeV LHC. The $p_T$ distribution of the leading Higgs bosons and $Z$ boson
are similar. Both peaks around 110 GeV. As would be expected, the leading Higgs boson's $p_T$
 is harder than next-to-leading Higgs boson which has $p_T$ distribution around 50 GeV with
 a significant tail. However, all three particles are produced mainly centrally. The leading
Higgs bosons and $Z$ boson are produced more back-to-back than the two
Higgs bosons which are produced relatively closer to each other. 
{ These features are also reflected in di-boson invariant mass distributions. 
For example, $M_{H_1Z}$ has harder tail than $M_{H_1H_2}$ and $M_{H_2Z}$. }
The partonic
centre-of-mass energy around 500 GeV contributes most to the cross section.
At higher center-of-mass energy, 100 TeV, the behavior of these distributions
are similar. We further compare the GG(LO) and QQ (LO and NLO) contributions 
in kinematic distributions for $p_T(H_1)$ and $p_T(Z)$ at 13 TeV LHC. In Fig.~\ref{fig:dists-qqhhz-gghhz-norm-sm} 
we see that contribution of GG channel is characteristically different from that of QQ channel.
The GG channel gives rise to softer events in $p_T(H_1)$, while 
harder events in $p_T(Z)$. These features remain true at higher centre-of-mass energies. 
In Fig.~\ref{fig:dists-qqhhz-gghhz-sm-13TeV}, we give the $p_T(H_1)$ and $p_T(Z)$ distributions 
combining QQ(NLO) and GG(LO) channels at 13  and 100 TeV. At the level of total cross section, the GG contribution is about 
14\% with respect to the QQ(NLO) contribution at 13 TeV. However, in the distributions the GG contribution 
can reach more than 20 \% in certain bins. Similarly, at 100 TeV  collider, although the contribution of the 
GG process to the total cross section is about $72\%$
of the QQ(NLO) contribution (see Table~\ref{table:xs-hhz}), in the case of $p_T(H_1)$ between $100-200$~GeV, the enhancement 
due to GG channel could be more than $100\%$.
 

\begin{figure}[!hbt]
\bc
\includegraphics [angle=0,width=0.45\linewidth]{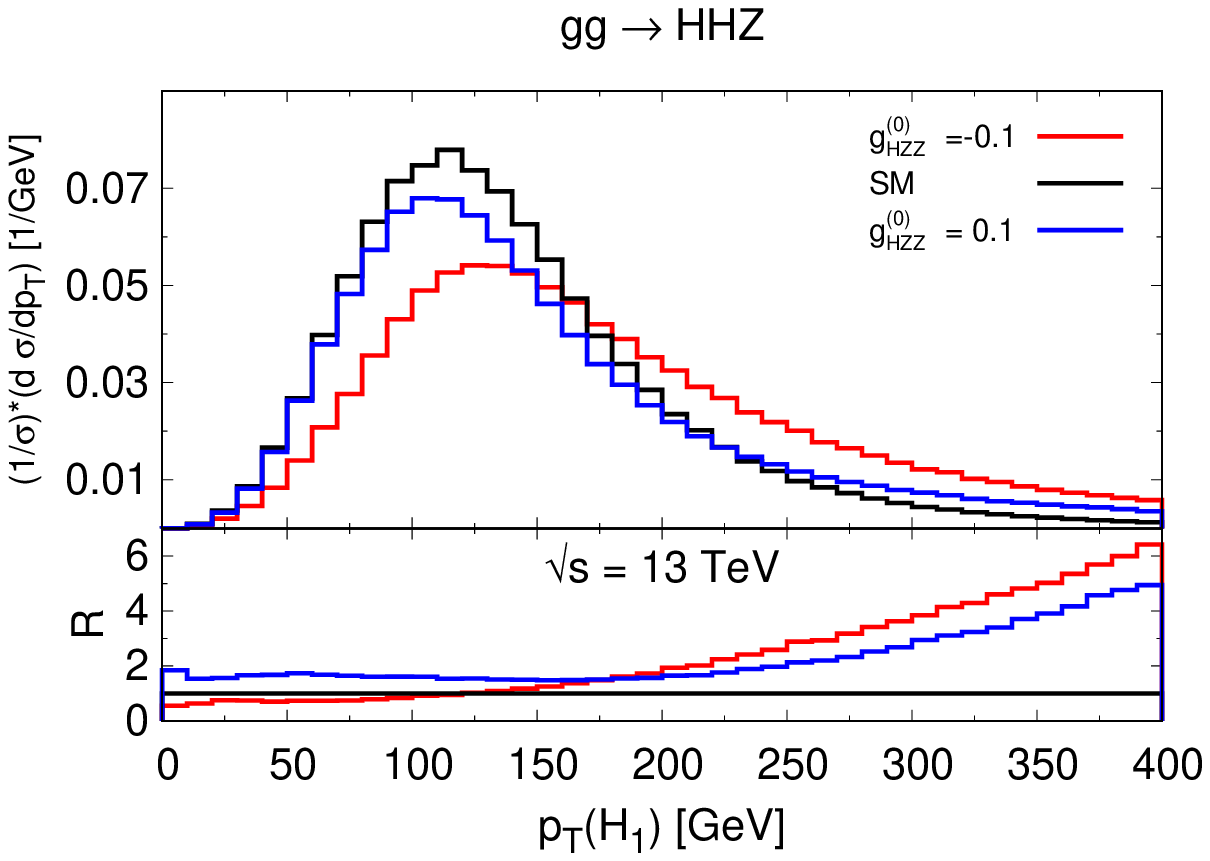}
\includegraphics [angle=0,width=0.45\linewidth]{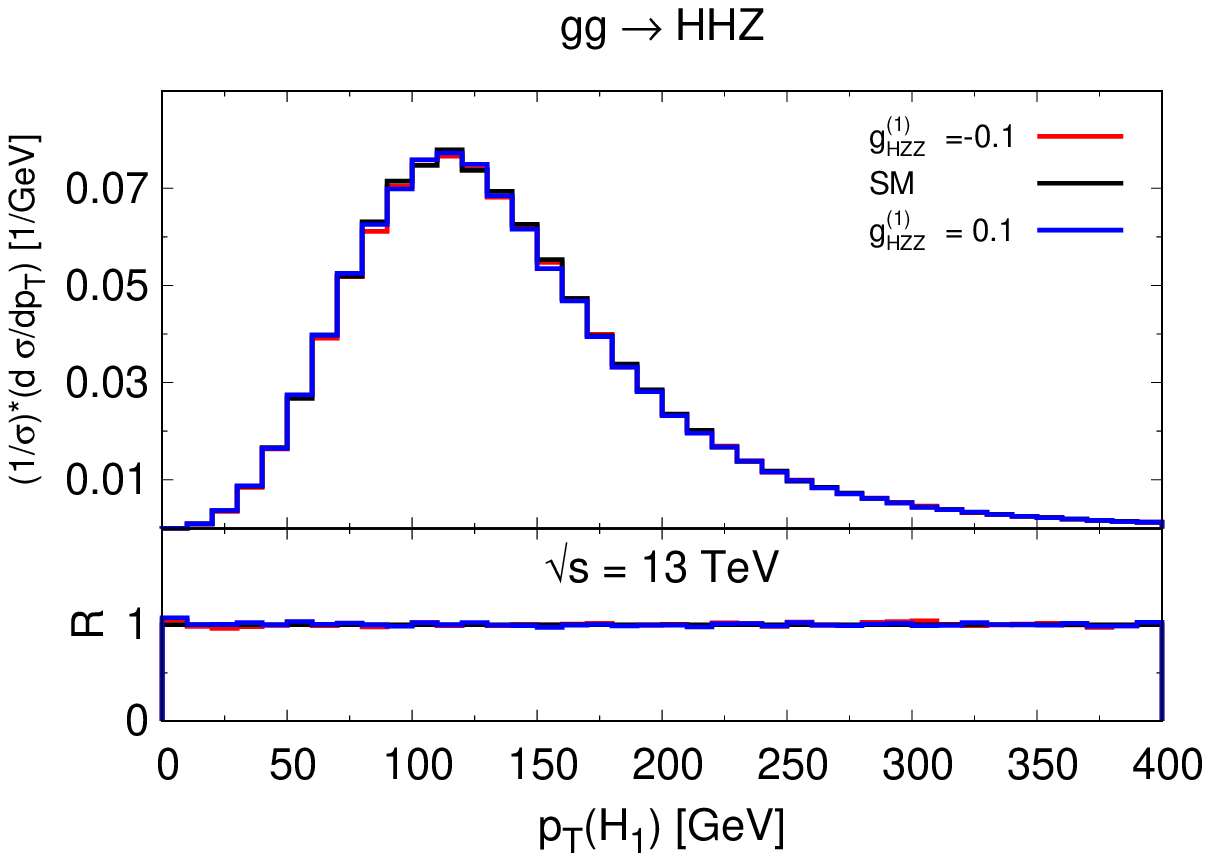}\\
\includegraphics [angle=0,width=0.45\linewidth]{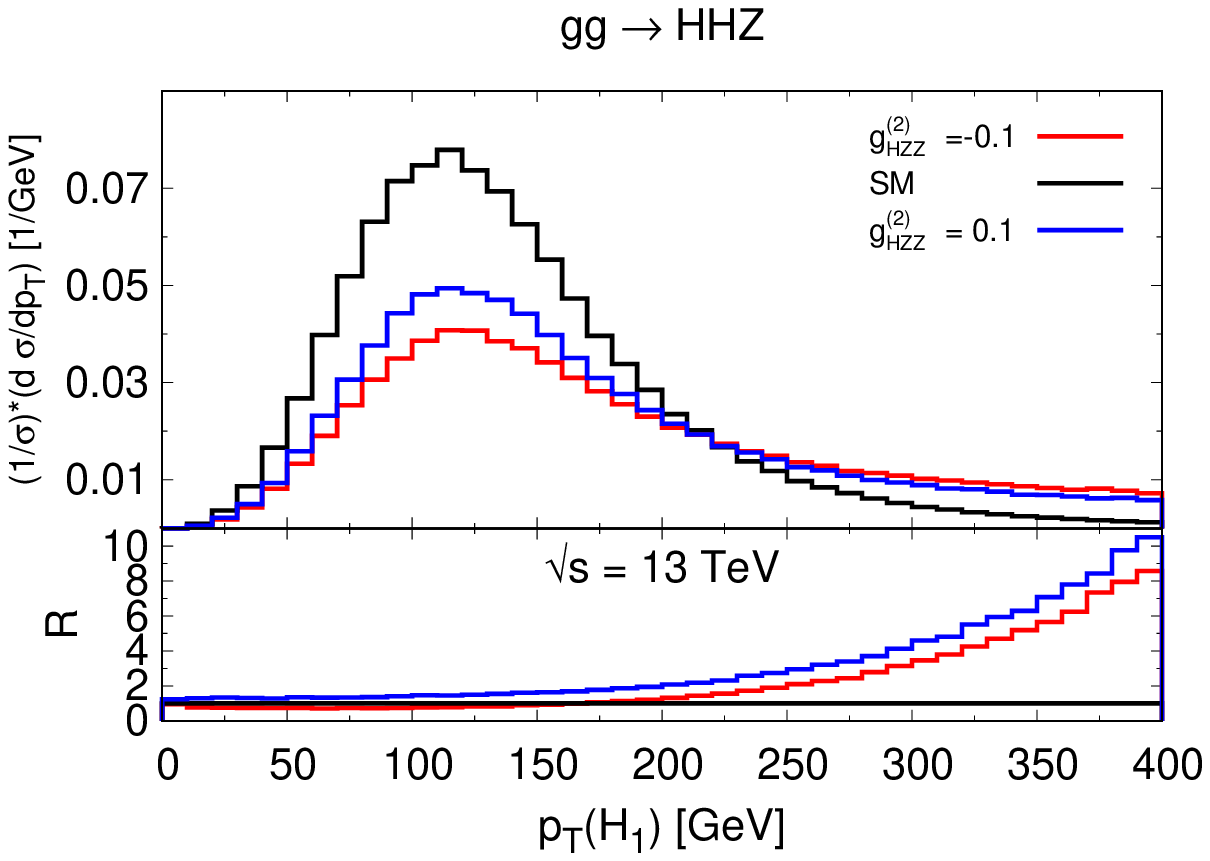}
\includegraphics [angle=0,width=0.45\linewidth]{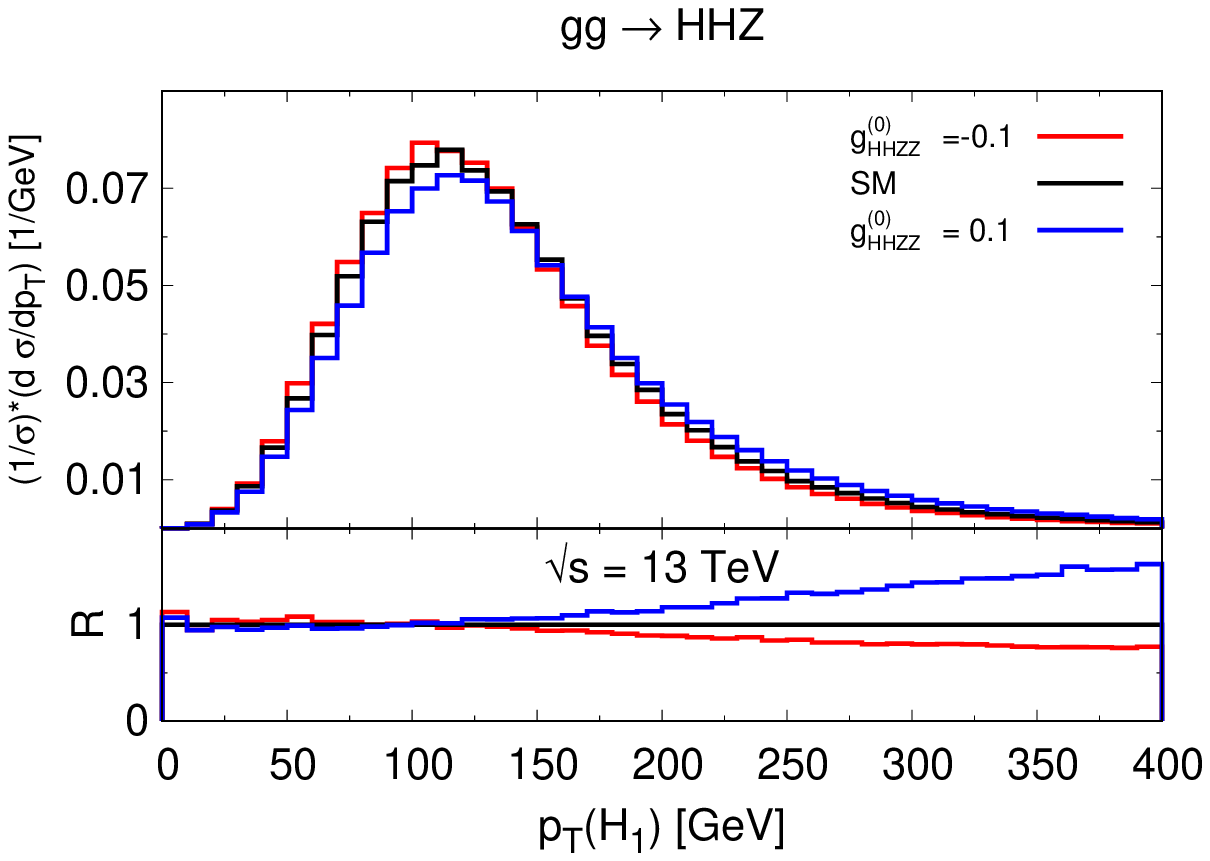}
\ec
\caption{ Normalized leading $p_T(H)$ distribution in $\GGHHZ$ at 13 TeV for some benchmark 
values of $HZZ$ and $HHZZ$ anomalous couplings. } \label{fig:dists-hhz-anml-hzz}
\end{figure}

  The process $gg \to HHZ$ has four different couplings -- $ttH, HZZ, HHH$, and $HHZZ$. In Fig.~\ref{fig:xs-hhz-anml},
 we probe sensitivity of the production cross section to these anomalous couplings.
The cross section is mainly sensitive to $ttH$ and $HZZ$ couplings. The scalar Yukawa coupling, $\kappa_\tin{t}$,   
and the scaling coupling, $g^{(0)}_{\tiny{HZZ}}$, lead to largest modification.
 The cross section could double with allowed parameter range of these parameters.  The effects of other anomalous couplings are rather modest.
 Like triple Higgs boson production, here also the $gg \to HHZ$ is not sensitive to ${\tilde \kappa}_\tin{t}$ in the allowed parameter range mentioned in section~\ref{sec:anml}.
 We find that tri-linear Higgs boson anomalous self-couplings do not play any significant role
 in this NNLO process. The derivative $HZZ$ couplings and $HHZZ$ coupling also don't play any significant role in the allowed range mentioned in section~\ref{sec:anml}.
 In Figs.~\ref{fig:dists-hhz-anml-yt},~\ref{fig:dists-hhz-anml-h3},~\ref{fig:dists-hhz-anml-hzz}, 
 the effects of various anomalous couplings relevant to $gg \to HHZ$ on leading $p_T(H)$ 
 distribution are shown. For that we take some benchmark values.
 We find that both ${\kappa}_\tin{t}$ and ${\tilde \kappa}_\tin{t}$ lead to harder tail compared to the SM prediction. 
 The contribution to the cross section at higher $p_T$ is significantly large for higher ${\kappa}_\tin{t}$.
 This may be a better avenue to see the effect of such interactions.
 Similar features are found for parameters $g^{(0)}_\tin{HZZ}$ and $g^{(2)}_\tin{HZZ}$. The effect 
 of $g^{(1)}_\tin{HZZ}$ is almost flat and remains close to SM prediction in all the bins.
 The dependence of the $p_T$ distribution on $g^{(0)}_\tin{HHZZ}$ is rather interesting. Compared to the SM values,
 this distribution is harder for the positive values of the anomalous coupling, and softer for the
 negative values. There seems to be harder tail of this $p_T$ distribution for
 larger values of $g^{(1)}_\tin{3H}$, specially for the positive values. The distribution for $g^{(0)}_\tin{3H}$ doesn't seem to show any special feature. The sensitivity of the NNLO process to anomalous interactions
 is similar at 100 TeV.

   In this paper our focus has been the NNLO $g g \to HHZ$ process. That is why we have
 presented detailed results for this process. One may ask how sensitive is LO process
 to anomalous interactions. We have explored this sensitivity by using {\tt Madgraph}. By including
 anomalous vertices, we find that LO quark-antiquark annihilation process is quite 
 sensitive to anomalous derivative $HZZ$ interaction. The cross section can increase
 by an order of magnitude. The increase is more at higher center-of-mass energy.
 This is unlike NNLO $gg \to HHZ$ process.

The process $pp \to HHZ$ is likely to be observable at the LHC. Including 
the NLO and NNLO corrections, the cross section is about 320 ab at 13 TeV. With the 
full integrated luminosity, one may expect around 1000 events. This process 
should be visible using various multilepton + jets signature. {The main irreducible
background of $ZZZ$ has the cross section of 9.2 fb.} But $Z \to b {\bar b}, \tau \tau$
branching ratios are smaller by a factor of 2-3 as compared to the Higgs boson
decay. Then by restricting the number of jets in the signature, one may be
able to detect this process. To look for evidence beyond the standard model,
one may look for enhancement in large $p_T(Z)$ events. Reducible backgrounds, 
as mentioned above, may be tamed by flavor-tagging of jets. At higher energy machines,
this process would definitely be observable.

\section{Conclusions}\label{sec:conclusions}

    In this paper, we have considered the processes -- $p p \to  HHH ,HH \gamma$, and $HHZ$. Our
    focus was on the gluon-gluon fusion contribution to them. The one-loop amplitude for the process
    $g g \to HH \gamma$ vanishes exactly. 
    The process $p p \to  HHH$ is important as it involves both trilinear and quartic Higgs boson couplings.
    A measurement of this process along with di-Higgs production can help in determining the form of the Higgs potential.
    It may be seen only if there exists
    anomalous interactions. This process is specially sensitive to trilinear Higgs boson couplings.
    This process can be observed at large center-of-mass energy machines with high luminosity. It will
    be challenging. 
    The process $pp \to HHZ$ may be observable at the LHC after accumulation of 3 ab$^{-1}$
    luminosity. 
    The GG(LO) contribution to this process 
    is actually a NNLO contribution 
    in $\alpha_s$, and due to a large gluon flux it is 14\% of the QQ(NLO) contribution to $p p \to  HHZ$ 
    at 13 TeV LHC. In certain kinematic windows GG(LO) contribution can be more than 20\%.
    At a 100 TeV machine, $\GGHHZ$ can be as important
    as $q {\bar q} \to HHZ$. This process is important, as it involves $HHH$ and $HHZZ$ couplings and is background to 
    triple Higgs production. The effect of $ttH$ and $HZZ$ anomalous couplings are more significant in the 
    distributions than in the total cross section.
    This process can definitely be observed at higher energy, such as 100 TeV, machines with enough
    luminosity.

\section*{Acknowledgement}
AS is supported by the MOVE-IN Louvain Cofund grant.

\appendix

\section{Feynman Rules for anomalous Higgs vertices}\label{sec:A-FR}

\vspace{0.2in}


The Feynman rules for various anomalous couplings of the Higgs boson 
considered in section~\ref{sec:anml} are given below:

\begin{center}
\begin{picture}(350,115)(20,-10)
\ArrowLine(0,90)(50,60)
\Text(-5,90)[r]{$t$}
\ArrowLine(50,60)(0,30)
\Text(-5,30)[r]{$\bar{t}$}
\linethickness{.02mm}
\multiput(50,60)(8,0){7}
{\line(1,0){4}}
\Text(117,60)[r]{$H$}
\Text(195,60)[l]{$-i\frac{m_\tin{t}}{v}\Big\{\left(1+ \kappa_\tin{t}\right) + i {\tilde \kappa}_\tin{t} \g_5\Big\}$}
\end{picture}
\end{center}


\begin{center}
\begin{picture}(350,115)(20,-10)
\DashArrowLine(0,60)(50,60){4}
\Text(-5,60)[r]{$H$}
\Text(30,70)[r]{$k_1$}
\DashArrowLine(90,90)(50,60){4}
\Text(107,90)[r]{$H$}
\Text(70,85)[r]{$k_2$}
\DashArrowLine(90,30)(50,60){4}
\Text(107,30)[r]{$H$}
\Text(70,35)[r]{$k_3$}
\Text(195,60)[l]{$-i{3m_\tin{H}^2\over v}
    \Big\{ (1 + g^{(0)}_{\text{\tiny 3}\tin{H}})\;  -
      {g^{(1)}_{\text{\tiny 3}\tin{H}}\over 3m_\tin{H}^2 }\; \sum^3_{j<k}\ p_j \cdot p_k \Big\}$}
\end{picture}
\end{center}

\begin{center}
\begin{picture}(350,115)(20,-10)
\DashArrowLine(0,60)(50,60){4}
\Text(-5,60)[r]{$H$}
\Text(30,70)[r]{$k_1$}
\DashArrowLine(90,90)(50,60){4}
\Text(107,90)[r]{$H$}
\Text(70,85)[r]{$k_2$}
\DashArrowLine(90,60)(50,60){4}
\Text(107,60)[r]{$H$}
\Text(85,65)[r]{$k_3$}
\DashArrowLine(90,30)(50,60){4}
\Text(107,30)[r]{$H$}
\Text(70,35)[r]{$k_4$}
\Text(195,60)[l]{$-i{3m_H^2\over v^2} \Big\{ (1 + g^{(0)}_{\text{\tiny 4}\tin{H}})  -
    {g^{(1)}_{\text{\tiny 4}\tin{H}}\over 6m_\tin{H}^2 }\sum^4_{j<k}\ p_j \cdot p_k \Big\}$}
\end{picture}
\end{center}

\begin{center}
\begin{picture}(350,115)(20,-10)
\DashArrowLine(0,60)(50,60){4}
\Text(-5,60)[r]{$H$}
\Photon(50,60)(90,90){2}{5}
\Text(107,90)[r]{$Z_\alpha$}
\ArrowLine(70,75)(65,72)
\Text(70,85)[r]{$k_1$}
\Photon(50,60)(90,30){2}{5}
\Text(107,30)[r]{$Z_\beta$}
\ArrowLine(70,45)(64,50)
\Text(70,35)[r]{$k_2$}
\Text(120,70)[l]{$i \dfrac{g M_\tin{Z}}{c_\tin{W}} \;
\Big\{ g^{\alpha \beta}(1+g^{(0)}_\tin{HZZ}) +
\dfrac{g^{(1)}_\tin{HZZ}}{M_\tin{Z}^2} \ [g^{\alpha \beta}(k_1 \cdot k_2) - k_2^\alpha
k_1^\beta] \;  + $}
\Text(195,35)[l]{\ \ \ \;\; $\dfrac{g^{(2)}_\tin{HZZ}}{M_\tin{Z}^2} \ [g^{\alpha \beta}(k_1^2 +k_2^2) -
(k_1^\alpha k_1^\beta + k_2^\alpha k_2^\beta )]
\Big\}$}
\end{picture}
\end{center}

\begin{center}
\begin{picture}(350,115)(20,-10)
\DashArrowLine(0,90)(50,60){4}
\Text(-5,90)[r]{$H$}
\DashArrowLine(0,30)(50,60){4}
\Text(-5,30)[r]{$H$}
\Photon(50,60)(90,90){2}{5}
\Text(107,90)[r]{$Z_\alpha$}
\Text(107,30)[r]{$Z_\beta$}
\Photon(50,60)(90,30){2}{5}
\ArrowLine(70,75)(65,72)
\ArrowLine(70,45)(64,50)
\Text(195,60)[l]{$ i \dfrac{g M_\tin{Z}}{c_\tin{W} v} \;
\Big\{ g^{\alpha \beta}(1+g^{(0)}_\tin{HHZZ}) \Big\}$}
\end{picture}
\end{center}

\begin{spacing}{1}

\end{spacing}
\clearpage

\end{document}